\documentclass[10pt,reqno]{article}
\usepackage{cite}
\usepackage[utf8]{inputenc}
\usepackage{rotating}
\usepackage[left=0.7in, right=1.0in, top=1.0in, bottom=1.0in]{geometry}
\usepackage{tabularx}
\usepackage{multirow}
\usepackage{booktabs}
\usepackage[list=true]{subcaption}
\usepackage{adjustbox}
\usepackage[graphicx]{realboxes}
\usepackage{amsmath}
\usepackage{color}
\usepackage{graphics}%
\usepackage{wrapfig}
\usepackage{eurosym}
\usepackage{pdflscape}

\usepackage{epstopdf}
\usepackage{graphicx}
\usepackage{placeins}
\usepackage{float}

\usepackage{fancyhdr}
 
\begin{document}
\title{\bf New solitary wave and Multiple soliton solutions\\ of (3 + 1)-dimensional KdV type equation\\ by using Lie symmetry approach
}

\author{Sachin Kumar$^a$\footnote{{\it sachinambariya@gmail.com}} and Dharmendra Kumar$^b$\footnote{{\it dharmendrakumar@sgtbkhalsa.du.ac.in}} \\ ${}^a$ \small Department of Mathematics, Faculty of Mathematical Scienes\\${}^b$ \small Department of Mathematics, SGTB Khalsa College,\\ \small University of Delhi, Delhi -110007, India} 
\maketitle
\begin{abstract}
Solitary waves are localized gravity waves that preserve their consistency and henceforth their visibility through properties of nonlinear hydrodynamics. Solitary waves have finite amplitude and spread with constant speed and constant shape.  In this paper, we have used Lie group of transformation method to solve (3 + 1)-dimensional KdV type equation. We have obtained the infinitesimal generators, commutator table of Lie algebra for the KdV type equation. We have achieved a number of exact solutions of KdV type equation in the explicit form through similarity reduction. All the reported results are expressed in analytic (closed form) and figured out graphically through their evolution solution profiles. We characterized the physical explanation of the obtained solutions with the free choice of the particular parameters by plotting some 3D and 2D illustrations. The geometrical analysis explains that the nature of solutions is travelling wave, kink wave, single solitons, doubly solitons and curve-shaped multisolitons.
\end{abstract}
\section{Introduction}
Solitons and nonlinear evolution equations (NLEEs) are broadly used to explain nonlinear phenomena in many mathematical physics and emerging engineering areas, such as nonlinear optics, condensed matter, plasma physics, fluid dynamics, convictive fluids, solid-state physics, acoustics, and quantum field theory \cite{whitham, kawahara, aspe,gara, cervero, depassier, selima}. 
Owing to the significant role  of solitons and nonlinear equations play in these scientific fields, constructing exact solutions for the NLEEs is of great value.

In the mid 19th century, John Scott Russell first investigated shallow water solitary waves experimentally and noted their importance through nonlinear interactions\cite{fiza,xliu}. 
Both Boussinesq and Rayleigh established mathematically the existence of steady solitary waves on shallow water, before Korteweg and de Vries (KdV) published their famous PDE, which was first originally derived by Boussinesq \cite{KdV1895, khattak, khuriKdV }. 
After this work, the theory of solitary waves remained almost untouched for 70 years until the mid 1960s when numerical studies by Zabusky and Kruskal \cite{zabusky} discovered the robust nature of soliton interactions, prompting an explosion of refined mathematical analysis on nonlinear PDEs. The history of solitary waves has been analyzed by Miles\cite{miles}, and that of water waves, more generally by Darrigol\cite{darrigol}. 

The Korteweg–de Vries (KdV) equation is given in the form of 
\begin{align}
u_t-6 u u_x+u_{xxx}=0.
\end{align}
The KdV type equation is characterized by the special 
waves which is called solitons on shallow water surfaces \cite{zabusky}. 
Solitons are localized wave disturbances that proliferate 
without changing shape or spreading out \cite{atom}. The KdV equation has a number of connections with physical problems such as shallow water waves with weakly non-linear restoring forces, ion acoustic waves in plasma, long internal waves in a density-stratified ocean, acoustic waves on a crystal lattice \cite{wiki}. 
The (3+1)-dimensional KdV type equation has many extensive applications in the field of condensed matter physics, fluid dynamics, plasma physics and optics \cite{sixing}.
 After the inverse scattering method, which is established for solving the KdV equation many important approaches were contributed by many researchers. In this view, we have observed many methods developed in the literature. For getting analytical solutions of considered system, homogeneous balance method, Hirota bilinear method, ansatz method, exp function method, Kudryashov simplest equation method and Lie group of transformation method can be cited \cite{hirota, wang, he, biswas, kudryashov,BookOlver, recio,ganda,laxmannan,STM,sachin1,sachin2,mukesh14,mukesh1,min,mawa3,wang2016,anjan2014,ray,sahoo}.

In this work, we study a (3+1)-dimensional KdV type equation of the form
\begin{align}\label{kdv}
\Delta := u_t+6 u_x u_y+u_{xxy}+u_{xxxxz}+60 u_x^2 u_z+10 u_{xxx} u_z+20 u_x u_{xxz}= 0
\end{align}
which was introduced by Lou \cite{lou} and obtained the 
five types of multidromion solutions for
its potentials form. Further, Wazwaz \cite{wazwaz2012} investigated one and two solitons solutions only. Besides this, the same problem was tackled by \"Unsal \cite{omer} and obtained complexiton and interaction solutions by Hirota Method.
This paper is a continuation to above mentioned earlier studies, in the sense that the (3+1)-dimensional KdV type equation is being analyzed using the Lie symmetry approach. The main objective of this paper is 
to obtain the symmetry reductions and exact solutions by using the Lie group of transformation method. 
We study various analytical (closed form) solutions of the equation via the symbolic calculations, including solitary waves solitons, single solitons, doubly solitons and multisolitons and travelling wave solitons. Furthermore, the exact solutions of the equation are graphically analyzed though their evolution profiles.
 
The skeleton of this paper is organized as follows: 
In Sec. 2, we perform the Lie group analysis
on the (3+1)-dimensional KdV type equation and present 
all the geometric vector fields. Then the complete symmetry classification of the (3+1)-dimensional KdV type Eq. \eqref{kdv} is performed. 
In Sec. 3, we discuss the Lie symmetry group of Eq. \eqref{kdv}. 
In Sec. 4, the symmetry reductions and exact solutions to the (3+1)-dimensional KdV type equation are investigated. 
In Sec. 5, discussion on graphical illustration of solutions are presented. 
Finally, the conclusion and some remarks are given in Sec. 6.

\section{Lie symmetry analysis for the (3+1) KdV type equation}
Lie group analysis is a powerful mathematical tool to study the 
properties of NLEEs and for obtaining the invariant solutions. 
If Eq \eqref{kdv} is invariant under a one-parameter 
Lie group of point transformations:
\begin{align}
\tilde{x} &=x+\epsilon\, \xi^1 (x,y,z,t,u) +O(\epsilon^2), \nonumber \\
\tilde{y} &=y+\epsilon\, \xi^2 (x,y,z,t,u) +O(\epsilon^2), \nonumber\\
\tilde{z} &=z+\epsilon\, \xi^3 (x,y,z,t,u) +O(\epsilon^2), \nonumber\\
\tilde{t} &=t+\epsilon\, \xi^4 (x,y,z,t,u) +O(\epsilon^2), \nonumber\\
\tilde{u} &=u+\epsilon\, \eta (x,y,z,t,u) +O(\epsilon^2), \nonumber
\end{align}
where $\epsilon$ is a small expansion parameter with infinitesimal generator
\begin{align}\label{eqv1v} 
{\bf V}=\xi^1(x,y,z,t,u) \frac{\partial}{\partial x}+ \xi^2(x,y,z,t,u) \frac{\partial}{\partial y}+\xi^3(x,y,z,t,u) \frac{\partial}{\partial z}+\xi^4(x,y,z,t,u) \frac{\partial}{\partial t}+\eta(x,y,z,t,u)\frac{\partial}{\partial u},\end{align}
then the vector field \eqref{eqv1v} generates a symmetry of Eq. \eqref{kdv}, and $V$ must satisfy Lie symmetry conditions
\begin{align}\label{pr}
pr^{(5)}V(\Delta)|_{\Delta=0} = 0,
\end{align}
where $pr^{(5)}V$ is the fifth prolongation of $V$.

Applying the fifth prolongation $pr^{(5)}V$ to Eq. \eqref{kdv}, the invariant conditions given by
\begin{align}
\eta^t+6 \eta^x u_y + 6 u_x \eta^y +\eta^{xxy}+\eta^{xxxxz}+&120 \eta^x u_x u_z 
+ 60 u_x^2 \eta^z+10 \eta^{xxx}u_z\notag\\
+&10 u_{xxx} \eta^z+20 \eta^x u_{xxz}+20 u_x \eta^{xxz} = 0,
\end{align}
where $\eta^t, \eta^x, \eta^y, \eta^z, \eta^{xxy}, \eta^{xxz}, \eta^{xxx}$ and $\eta^{xxxxz}$ are the coefficients of $pr^{(5)}V(\Delta)$. 
Moreover, we have
\begin{align}\label{coeff}
\eta^t &= D_t(\eta)-u_x D_t (\xi^1)-u_y D_t (\xi^2)-u_z D_t (\xi^3)-u_t D_t (\xi^4),\notag\\
\eta^x &= D_x(\eta)-u_x D_x (\xi^1)-u_y D_x (\xi^2)-u_z D_x (\xi^3)-u_t D_x (\xi^4),\notag\\
\eta^y &= D_y(\eta)-u_x D_y (\xi^1)-u_y D_y (\xi^2)-u_z D_y (\xi^3)-u_t D_y (\xi^4),\notag\\
\eta^z &= D_z(\eta)-u_x D_z (\xi^1)-u_y D_z (\xi^2)-u_z D_z (\xi^3)-u_t D_z (\xi^4),\notag\\
\eta^{xxx} &= D_x (\eta_{xx})-u_{xxx} D_x (\xi^1)- u_{xxy} D_x (\xi^2)-u_{xxz} D_x (\xi^3)-u_{xxt} D_x (\xi^4),\notag\\
\eta^{xxy} &= D_y (\eta_{xx})-u_{xxx} D_y (\xi^1)- u_{xxy} D_y (\xi^2)-u_{xxz} D_y (\xi^3)-u_{xxt} D_y (\xi^4),\notag\\
\eta^{xxz} &= D_z (\eta_{xx})-u_{xxx} D_z (\xi^1)- u_{xxy} D_z (\xi^2)-u_{xxz} D_z (\xi^3)-u_{xxt} D_z (\xi^4),\notag\\
\eta^{xxxxz} &= D_z (\eta_{xxxx})-u_{xxxxx} D_z (\xi^1)- u_{xxxxy} D_z (\xi^2)-u_{xxxxz} D_z (\xi^3)-u_{xxxxt} D_z (\xi^4),
\end{align}
where $D_x$, $D_y$ and $D_t$ represent the 
total derivative operators for $x$, $y$ and $t$, respectively. 
For example, one of them can be given as
\begin{align}
D_x = \frac{\partial}{\partial x}+ u_x\frac{\partial}{\partial u}+u_{xx}\frac{\partial}{\partial u_x}+u_{xy}\frac{\partial}{\partial u_y}+u_{xt}\frac{\partial}{\partial u_t}+\dots. 
\end{align}
In similar manner, we can use the total derivative operators for other variables.
Incorporating all the expressions of Eq. \eqref{coeff} into Eq. \eqref{pr} and then comparing the various differential coefficients of $u$ to zero,
we get the following determining system of equations:
\begin{align}\label{deteq}
\eta_t=0,\,\,\, 2 \eta_u = -\xi^4_t+\xi^2_y, \,\,\,
6 \eta_x= \xi^2_t,\,\,\, 6 \eta_y = \xi^1_t,\notag\\
10 \eta_z =\xi^1_y,\,\,\, \xi^4_u=\xi^4_x=\xi^4_y=\xi^4_z=\xi^4_{tt}=0,\notag\\
\xi^1_u=0,\,\,\, 2 \xi^1_x =\xi^4_t-\xi^2_y,\,\,\, \xi^1_z=\xi^1_{tt}=\xi^1_{ty}=\xi^1_{yy}=0,\notag \\
\xi^2_u=\xi^2_x= \xi^2_z=\xi^2_{tt}=\xi^2_{ty}=\xi^2_{yy}=0,\notag\\
\xi^3_t=\xi^3_u=\xi^3_x=0,\,\,\, 3 \xi^3_y= 10 \xi^2_t,\,\,\, \xi^3_z=-\xi^4_t+2\xi^2_y,
\end{align}
where $\eta_t=\frac{\partial \eta}{\partial t}, \eta_x=\frac{\partial \eta}{\partial x},
\eta_u=\frac{\partial \eta}{\partial u},
\xi^1_x=\frac{\partial \xi^1}{\partial x}, \xi^4_{tt}=\frac{\partial^2 \xi^4}{\partial t^2},
\xi^1_{ty}=\frac{\partial^2 \xi^1}{\partial t \partial y}, etc$.  

Solutions of Eq. \eqref{deteq} resulted in the following infinitesimal generators:
\begin{align}\label{gener}
\xi^1  &= \frac{1}{4} (a_1-a_5)x + a_8 t+ a_9 y +a_{10}, \notag\,\, \\
\xi^2 &= \frac{1}{2} (a_1 +a_5)y+\frac{3t}{10} a_4+a_7,\,\,\notag \\
\xi^3 &= a_4 y+ a_5 z + a_6,\notag \\
\xi^4&=a_1 t+a_3,\notag \\
\eta&=\frac{1}{4}(a_5-a_1)u+\frac{x}{20} a_4+\frac{y}{6}a_8 +\frac{z}{10}a_9  +a_2,
\end{align}
where $a_i, i=1,\dots, 10$ are all integral constants. 
For the  above cumbersome calculations, computer software Maple is used.
Therefore, Lie algebra of infinitesimal symmetries of Eq. \eqref{kdv} is spanned by the following ten vector fields:
\begin{align}\label{vec}
v_1=&\frac{x}{4}\frac{\partial}{\partial x}
+\frac{y}{2}\frac{\partial}{\partial y}
+t\frac{\partial}{\partial t}
- \frac{u}{4} \frac{\partial}{\partial u},  \,\,\,\,\,\,\,\,\,\,\,\,\,\,\,\,\,\,
v_2= \frac{\partial}{\partial u},  \,\,\,\,\,\,
v_3=
\frac{\partial}{\partial t},  \,\,\,\,\,\,
v_4= 
\frac{3t}{10}\frac{\partial}{\partial y}
+y \frac{\partial}{\partial z}
+\frac{x}{20} \frac{\partial}{\partial u}, \notag \\ \,\,\, \,\,\, 
v_5=&
\frac{-x}{4}\frac{\partial}{\partial x}
+\frac{y}{2} \frac{\partial}{\partial y}
+z \frac{\partial}{\partial z}
+\frac{u}{4} \frac{\partial}{\partial u},  \,\,\,\,\,\,\,\,\,\,\,\,
v_6= \frac{\partial}{\partial z}, \,\,\,\,\,\,  
v_7= \frac{\partial}{\partial y},\notag\\  \,\,\,\,\,\,
v_8=& t\frac{\partial}{\partial x}+ \frac{y}{6}\frac{\partial}{\partial u},\,\,\,\,\,\,\,\,\,\,\,\,\,\,\,\,\,\,\,\,\,\,\,\,\,\,\,\,\,\,\,\,\,\,\,\,\,\,\,\,\,\,\,\,\,\,\,\,\,\,\,\,\,\,\,\,\,
v_9= y\frac{\partial}{\partial x}+\frac{z}{10}\frac{\partial}{\partial u},\,\,\,\,\,\,\,\,\,\,\,\,
v_{10}= \frac{\partial}{\partial x}.  
\end{align}

\begin{table}
\caption{Commutation table of Lie algebra}\centering
\begin{tabular}{ccccccccccc}
\hline
*   & $v_1$ & $v_2$ & $v_3$ & $v_4$ & $v_5$ & $v_6$ & $v_7$ & $v_8$& $v_9$ & $v_{10}$ \\
\hline
$v_1$ & 0   & $\frac{1}{4} v_2$  & $-v_3$ & $\frac{1}{2}v_4$  & 0 & 0 & $-\frac{1}{2} v_7$  & $\frac{3}{4} v_8$ & $ \frac{1}{4} v_9$ & $\frac{-1}{4}v_{10}$ \\
$v_2$ & -$\frac{1}{4}v_2$   & 0  &0 & 0   & $\frac{1}{4} v_2$  & 0 &0  & 0&0&0\\
$v_3$ & $v_3$   & 0  & 0 & $\frac{3}{10}v_7$   & 0  & 0 & 0  & $v_{10}$ &0&0\\
$v_4$ & -$\frac{1}{2}v_4$   & 0  &$-\frac{3}{10}v_7$ & 0 & $\frac{1}{2}v_4$& 0& $-v_6$& 0&$\frac{3}{10}v_8$ &$-\frac{1}{20}v_2$\\
$v_5$ & 0  & $-\frac{1}{4}v_2$  & 0 & -$\frac{1}{2}v_4$ & 0& $-v_6$& $-\frac{1}{2}v_7$& $\frac{1}{4} v_8$& $\frac{3}{4}v_9$& $\frac{1}{4} v_{10}$\\
$v_6$ & 0  &  0  & 0 & 0 & $v_6$& 0& 0& 0 &$\frac{1}{10}v_2$   &0\\
$v_7$ & $\frac{1}{2} v_7$  &  0  & 0 & $v_6$ & $\frac{1}{2}v_7$& 0& 0& $\frac{1}{6}v_2$& $v_{10}$ & 0\\
$v_8$ & -$\frac{3}{4} v_8$  & 0  &   $-v_{10}$ & 0& $-\frac{1}{4}v_{8}$& 0& $-\frac{1}{6}v_2$& 0&0&0\\
$v_9$ & $-\frac{1}{4}v_9$  		  & 0  &   0 & $-\frac{3}{10}v_8$& $-\frac{3}{4}v_9$& $-\frac{1}{10} v_2$& $-v_{10}$& 0&0&0 \\
$v_{10}$ & $\frac{1}{4}v_{10}$  & 0  &   0 & $\frac{1}{20}v_2$ & $-\frac{1}{4}v_{10}$ & 0 & 0 & 0 & 0 & 0\\
\hline
\end{tabular}
\label{tab1}
\end{table}
It shows that the symmetry generators found in Eqs. \eqref{vec} form the 10 - dimensional Lie algebra. Also, it is easy to check that the symmetry generators found in \eqref{vec} form a closed Lie algebra whose commutation relations are given in Table \ref{tab1}. Then, all of the infinitesimal of Eq. \eqref{kdv} can be expressed as a linear combination of $v_i$ given as 
\begin{align*}
{\bf V}=a_1 v_1+a_2 v_2+a_3 v_3+a_4 v_4+a_5 v_5+a_6 v_6+ a_7 v_7+ a_8 v_8+a_9v_9+a_{10} v_{10}.
\end{align*}
The $(i, j)$th entry of the Table \ref{tab1} is the Lie bracket $[v_i \,\, v_j] = v_i \cdot v_j-v_j \cdot v_i$.  
We observe that Table \ref{tab1} is skew-symmetric with zero diagonal elements.
Also, Table \ref{tab1} shows that the generators $v_i,\,\, 1 \le i \le 10$ are linearly independent. 
\section{Symmetry group of (3+1)- dimensional KdV type equation}
In this section, we obtain the group transformations 
\begin{align}
g_i:(x,y,z,t,u) \rightarrow (\tilde{x}, \tilde{y}, \tilde{z}, \tilde{t}, \tilde{u}),
\end{align}
which is generated by the generators of infinitesimal transformations $v_i$ for $1 \le i \le 10$.  In order to get some exact solutions from known ones, we should find the Lie symmetry groups from the related symmetries. To get the Lie symmetry group, we should solve the following problems
For this purpose, we need to solve, following system of ODE's
\begin{align}
\frac{d}{d \epsilon} (\tilde{x}, \tilde{y}, \tilde{z}, \tilde{t}, \tilde{u})&={\bf \sigma}(\tilde{x}, \tilde{y}, \tilde{z}, \tilde{t}, \tilde{u}), \\
(\tilde{x}, \tilde{y}, \tilde{z}, \tilde{t}, \tilde{u})|_{\epsilon=0} &= (x,y,z,t,u),
\end{align}
where $\epsilon$ is an arbitrary real parameter and 
\begin{align}
{\bf \sigma}=\xi^1 u_x+\xi^2 u_y+\xi^3 u_z+\xi^4 u_t+\eta u.
\end{align}
So, we can obtain the Lie symmetry group
\begin{align}
g:(x,y,z,t,u) \rightarrow (\tilde{x}, \tilde{y}, \tilde{z}, \tilde{t}, \tilde{u}).
\end{align}
According to different $\xi^1, \xi^2, \xi^3, \xi^4$, and $\eta$, we have the following groups
\begin{align}
g_1:&(x,y,z,t,u) \rightarrow(x e^{\epsilon }, y e^{\epsilon }, z, t e^{4 \epsilon }, u e^{-\epsilon }),\notag \\
g_2:&(x,y,z,t,u) \rightarrow(x,y,z,t,u+\epsilon),\notag \\
g_3:&(x,y,z,t,u) \rightarrow(x,y,z,t+\epsilon,u),\notag \\
g_4:&(x,y,z,t,u) \rightarrow( x, y + 6 t \epsilon, z + 60 t \epsilon ^2+20 y \epsilon, t, u+\epsilon x ),\notag \\
g_5:&(x,y,z,t,u) \rightarrow(x e^{-\epsilon }, y e^{2 \epsilon }, z e^{4 \epsilon }, t, u e^{\epsilon }),\notag \\
g_6:&(x,y,z,t,u) \rightarrow(x,y,z+\epsilon,t,u),\notag \\
g_7:&(x,y,z,t,u) \rightarrow(x,y+\epsilon,z,t,u),\notag \\
g_8:&(x,y,z,t,u) \rightarrow(x+6 \epsilon t ,y,z,t,u+ \epsilon y),\notag \\
g_9:&(x,y,z,t,u) \rightarrow(x+10 \epsilon y,y,z,t,u+\epsilon z),\notag \\
g_{10}:&(x,y,z,t,u) \rightarrow(x+\epsilon,y,z,t,u).\notag
\end{align}
The entries on the right side give the transformed point $\exp(x,y,z,t,u)=(\tilde{x}, \tilde{y}, \tilde{z}, \tilde{t}, \tilde{u})$. 
The symmetry groups $g_2, g_3,  g_6, g_7$ and $g_{10}$ demonstrate the
space and time invariance of the equation. 
The well known scaling symmetry turns up in $g_1, g_4, g_5, g_8$ and $g_9$. We can obtain the corresponding
new solutions by applying above groups $g_i, 1 \le i \le 10$.

If $u=f(x,y,z,t)$ is a known solution of Eq. \eqref{kdv}, then by using above groups $g_i, 1 \le i \le 10$ corresponding new solutions $u_i, 1 \le i \le 10$  can be obtained as follows
\begin{align}
u_1 &= e^{\epsilon } f_1(x e^{-\epsilon }, y e^{-\epsilon }, z, t e^{-4 \epsilon }), \notag \\
u_2 &= f_2(x,y,z,t)-\epsilon, \notag \\
u_3 &= f_3(x,y,z,t-\epsilon), \notag \\
u_4 &= f_4( x, y - 6 t \epsilon, z - 60 t \epsilon ^2 - 20 y \epsilon, t)- \epsilon x, \notag \\
u_5 &= e^{-\epsilon }f_5(x e^{\epsilon }, y e^{-2 \epsilon }, z e^{-4 \epsilon }, t), \notag \\
u_6 &= f_6(x,y,z-\epsilon,t), \notag \\
u_7 &= f_7(x,y-\epsilon,z,t), \notag \\
u_8 &= f_8(x-6 t \epsilon,y,z,t)-\epsilon y, \notag \\
u_9 &= f_9(x-10 \epsilon y,y,z,t)-\epsilon z, \notag \\
u_{10} &= f_{10}(x-\epsilon,y,z,t), 
\end{align}
By selecting the arbitrary constants, one can obtain many new solutions.
Thus, to obtain the invariant solutions of Eq. \eqref{kdv}, the corresponding Lagrange system is
\begin{align*}
\frac{dx}{\xi^1(x,y,z,t)}=\frac{dy}{\xi^2(x,y,z,t)}=\frac{dz}{\xi^3(x,y,z,t)}=\frac{dt}{\xi^4(x,y,z,t)}=\frac{du}{\eta(x,y,z,t)}.
\end{align*}
The different forms of the invariant solutions of the equation are obtained by assigning the specific values to $a_i,\,\, 1 \le i \le 10$. Therefore, the Lie symmetry method predicts the following vector fields to generate the different forms of the invariant solutions.

\section{Symmetry reduction and closed-form solutions of (3+1) KdV type equation}
In this section, we systematically derive the Lie point symmetries of the (3+1)-dimensional KdV type equation. For this purpose, we reduce the characteristic equations of vector fields obtained in the previous section for getting reduction equations.
\subsection{\noindent \textbf{\textit{Vector field $v_1$:}}}
For the infinitesimal generator
\begin{eqnarray}
v_1 = \frac{x}{4}\frac{\partial}{\partial x}
+\frac{y}{2}\frac{\partial}{\partial y}
+t\frac{\partial}{\partial t}
- \frac{u}{4} \frac{\partial}{\partial u}, 
\end{eqnarray}
the characteristic equation is given by
\begin{align}
\frac{dx}{\frac{x}{4}}=\frac{{dy}}{\frac{y}{2}}=\frac{{dz}}{0}=\frac{{dt}}{t}=\frac{{du}}{\frac{u}{4}}.
\end{align}
Then, similarity form of the KdV type equation \eqref{kdv} yield
\begin{align}\label{v1uF}
u=F(X,Y,Z),
\end{align}
where $F(X,Y,Z)$ is similarity function in which similarity variables $X, Y$ and $Z$ can be expressed as
\begin{align}
X=x t^{-\frac{1}{4}}, Y=y t^{-\frac{1}{2}}, Z=z.
\end{align}
Using Eq. \eqref{v1uF} into Eq. \eqref{kdv}, we obtain the following (2+1)- dimensional nonlinear
PDE with variable coefficients as first reduction of the equation given as
\begin{align}\label{v1inF}
F-4 \left(10 F_Z \left(6 F_X^2+F_{\text{XXX}}\right)+20 F_X F_{\text{XXZ}}+F_{\text{XXXXZ}}+F_{\text{XXY}}\right)+2 F_Y \left(Y-12 F_X\right)+X F_X=0,
\end{align}
where $F_X = \frac{\partial F}{\partial X}, F_Y = \frac{\partial F}{\partial Y}$, etc.
To solve Eq. \eqref{v1inF}, we obtain new set of infinitesimal 
for $X, Y, Z$ and $F$ by applying similarity transformations 
method (STM) which are given below
\begin{align}\label{v1gen}
\xi_X =-\frac{X}{4}A_1,\,\,
\xi_Y =\frac{Y}{2} A_1,\,\,
\xi_Z =A_1 Z +A_2,\,\,
\eta_F = A_1 \frac{F}{4},
\end{align}
where $A_1$ and $A_2$ are arbitrary constants.

Consequently, the different aspects of integral constants, this case can be split into the following two subcases.
\subsubsection{For $A_1 \ne 0, A_2 = 0$ in Eq. \eqref{v1gen}}

In this subcase, we have following characteristic equation
\begin{align}\label{v1A1cheq}
\frac{dX}{-\frac{X}{4}}=\frac{{dY}}{\frac{Y}{2}}=\frac{{dZ}}{Z}=\frac{dF}{\frac{F}{4}}.
\end{align}
Further, the function $F$ can be written in new similarity form as 
\begin{align}
F=Z^{\frac{1}{4}} G(r,s), 
\end{align}
where
$r = X Z^{\frac{1}{4}}$ and  $s= \frac{Y}{\sqrt{Z}}$ are similarity variables. 
By substituting similarity form in Eq. \eqref{v1inF}, we have following reduced (1+1)-dimensional nonlinear partial differential equation which is given as follows:
\begin{align}\label{v1a1Grs}
60 r G_r ^3+G ( 10 G_{rrr}+60G_r^2-1 ) +&G_r (24G_s+60G_{rr}-40s G_{rrs}+r(30 G_{rrr}))+4G_{rrs}\notag\\
+&5G_{rrrr}+rG_{rrrrr}-2 s(G_s (1+60 G_r^2+10G_{rrr})+G_{rrrrs})  =0
\end{align}
We could not find generators of Eq. \eqref{v1a1Grs} because of high non-linearity. Hence, this equation can be solved numerically.
\subsubsection{For $A_1 = 0, A_2 \ne 0$ in Eq. \eqref{v1gen} }

We have following characteristic equation
\begin{align}\label{v1A1}
\frac{dX}{0}=\frac{{dY}}{0}=\frac{{dZ}}{1}=\frac{dF}{0}.
\end{align}
By solving Eq. \eqref{v1A1}, we obtain the similarity variable as follows
\begin{align}\label{eq27}
r = X\,\,\,  \text{and}\,\,\,\,  s= Y \,\,\, \text{ and the similarity form as}\,\,\,\,\, F=G(r,s).
\end{align} 
By substituting group invariant solution in Eq. \eqref{v1inF}, we have following reduced (1+1)- dimensional nonlinear PDE which is given as follows:
\begin{align}\label{v1a2Grs}
G+2 G_s(s-12 G_r)+r G_r - 4 G_{rrs}=0
\end{align}
Again, we can find infinitesimals for Eq. \eqref{v1a2Grs} given as
\begin{align}\label{v1a2}
\xi_r =-\frac{r}{2} b_1,\,\,
\xi_s =b_1 s,\,\,
\xi_G =\frac{G}{2}b_1.
\end{align}
where $b_1$ is an arbitrary constant. Then established characteristic equation for Eq. \eqref{v1a2} is
\begin{align}
\frac{dr}{-\frac{r}{2} b_1}=\frac{ds}{b_1 s}=\frac{dG}{\frac{G}{2}b_1}.
\end{align}
We obtain the similarity variable as 
$w = r \sqrt{s}$
and the similarity form is given by
\begin{align}\label{eq30}
G(r,s)=\frac{1}{r} R(w).
\end{align} 
Substituting the value of $G$ in Eq. \eqref{v1a2Grs} we obtain following nonlinear ODE given as
\begin{align}\label{v1inR}
w^2 R^{(3)}-R' \left(-6 w R'+6 R+w^2\right)=0,
\end{align}
where $'$ denotes the derivative with respect to $w$.
Eq. \eqref{v1inR} is a complicated nonlinear ODE and cannot be solved in general. Anyhow, assuming the adequate values of arbitrary constants, some particular results are given below
\begin{align}\label{eq32}
R(w)=\alpha_1, \,\,\,R(w)=\frac{1}{6} w^2,
\end{align}
where $\alpha_1$ is an arbitrary constant.
Using Eqs. \eqref{eq32}, \eqref{eq30}, \eqref{eq27} in Eq. \eqref{v1uF}, 
we obtain algebraic rational solutions of Eq. \eqref{kdv} as
\begin{align}
u_1(x,y,z,t)=\frac{\alpha_1}{x},\,\,\,
u_2(x,y,z,t)=\frac{x  y}{6 t}.
\end{align}

\subsection{\noindent \textbf{\textit{Vector field $v_2$:}}}
The associated Lagrange system for
\begin{align}
v_2=\frac{\partial}{\partial u} \label{eqv2}
\end{align}
is given by
\begin{align}
\frac{dx}{0}=\frac{{dy}}{0}=\frac{{dz}}{0}=\frac{{dt}}{0}=\frac{{du}}{1}
\end{align}
The nontrivial solution in this case cannot be obtained.
\subsection{\noindent \textbf{\textit{Vector field $v_3$:}}} 
The associated Lagrange system for 
\begin{align*}
v_3=\frac{\partial}{\partial t}
\end{align*}
is given by
\begin{align}
\frac{dx}{0}=\frac{{dy}}{0}=\frac{{dz}}{0}=\frac{{dt}}{1}=\frac{{du}}{0}.
\end{align}
Similarity reduction of Eq. \eqref{kdv} is
\begin{align}\label{v3uF}
u=F(X,Y,Z),
\end{align}
where $X=x, Y=y, Z=z$ 
are the three invariants that we obtained.
Substituting Eq. \eqref{v3uF} into Eq. \eqref{kdv}, we obtain the following nonlinear
PDE with three independent variables
\begin{align}\label{v3inF}
6F_X F_Y + 20 F_X F_{XXZ} +F_{XXY} +10F_Z (6F_X^2+ F_{XXX}) +F_{XXXXZ}=0,
\end{align}
we obtain solution of Eq. \eqref{v3inF} as
\begin{figure}[h!]
\centering
\subcaptionbox{$\forall \,y$}{\includegraphics[width=0.300\textwidth]{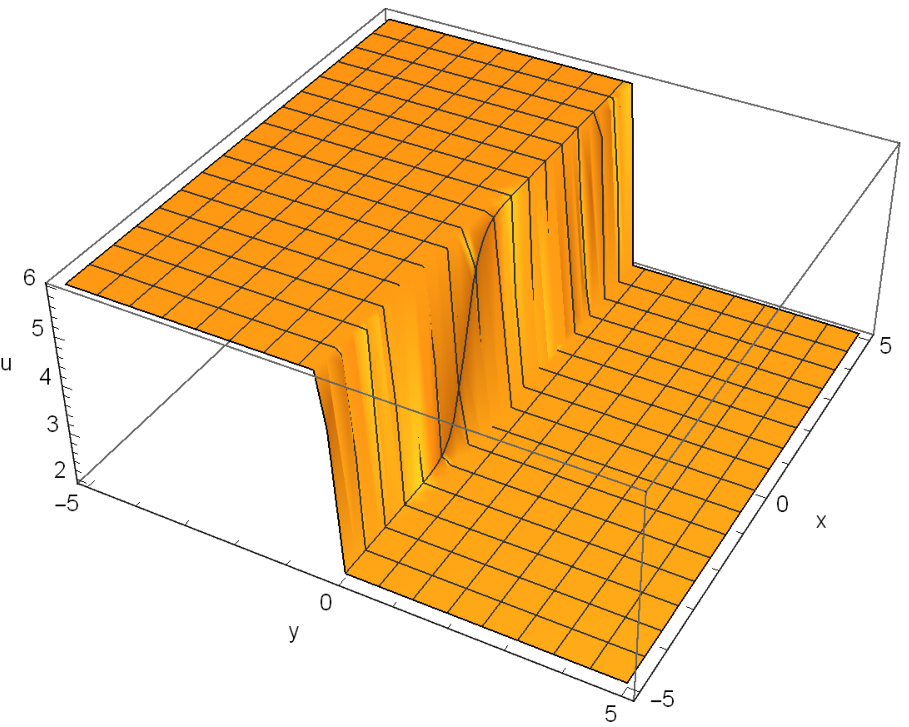}}%
\hfill
\subcaptionbox{$x\,y$ -plane}{\includegraphics[width=0.35\textwidth]{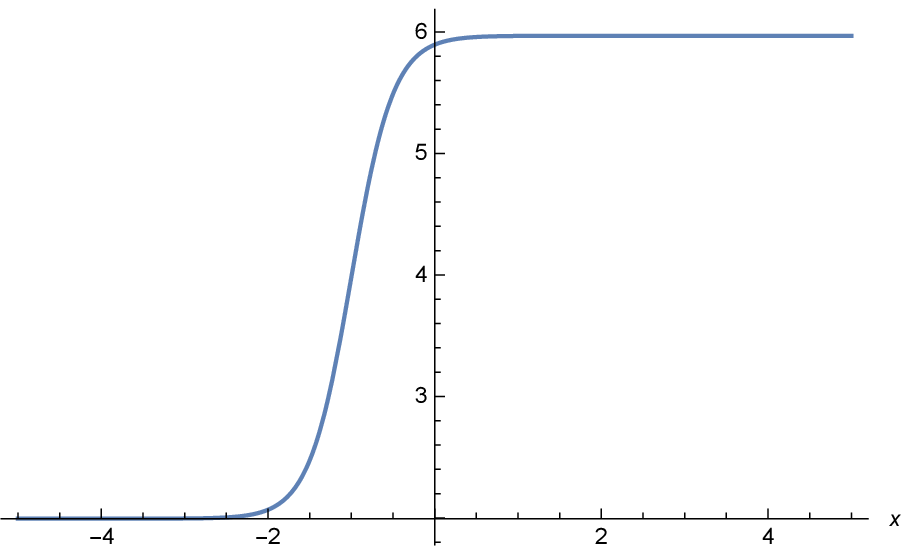}}%
\hfill 
\subcaptionbox{$\forall \,y$}{\includegraphics[width=0.20\textwidth]{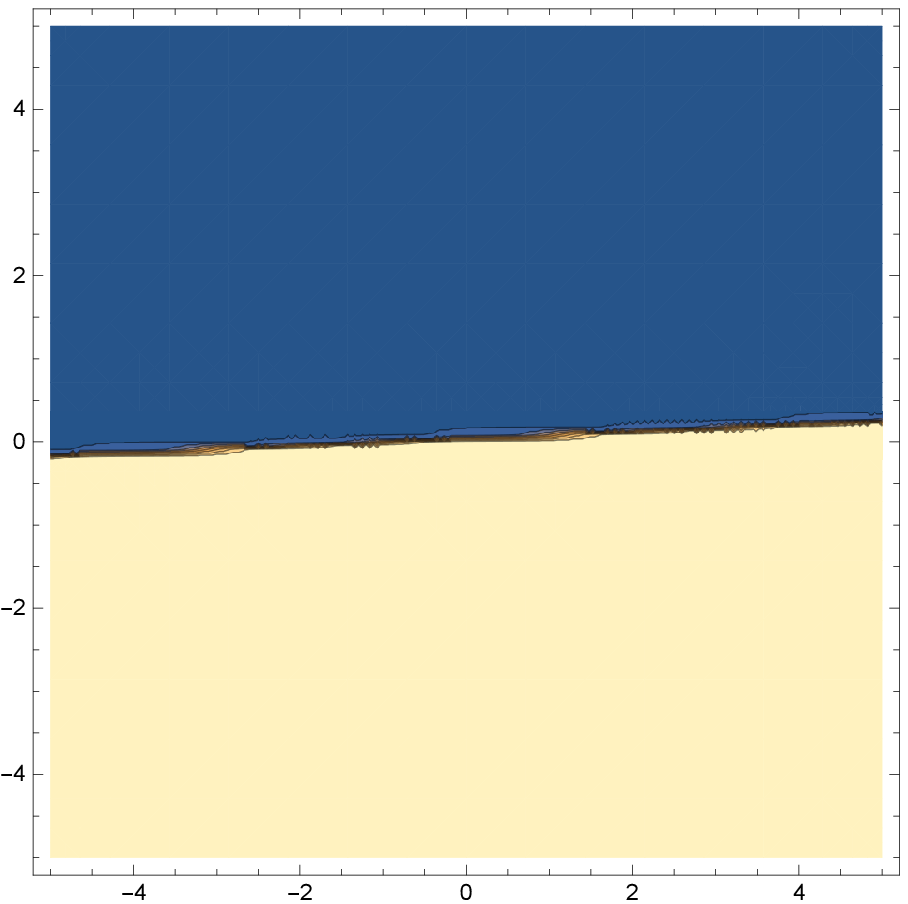}}%
\hfill 
\caption{Travelling wave profile in $x\,y$--plane of Eq. \eqref{v3A3} for the parameters $c_1 = 1.9872, c_3 = 2.9876, c_4 = 1.9876, c_5 = 3.9812.$ 
(a) Strip soliton. 
(b) The wave propagation pattern of the wave along the $x$-axis. 
(c) Contour plot. }
\label{f1v3A4}
\end{figure}

\begin{align}\label{v3A3sol1}
F(X,Y,Z)&=c_1 \tanh \left(c_1 X-4 c_3 c_1^2 Y+c_3 Z+c_4\right)+c_5.
\end{align}
Hence hyperbolic function solution of Eq. \eqref{kdv} is given as
\begin{align}
u_3(x,y,z,t)&=c_1 \tanh \left(c_1 x-4 c_3 c_1^2 y+c_3 z+c_4\right)+c_5,\label{v3A3}
\end{align}
where  $c_1, c_2, c_3, c_4$,and $c_5$ are arbitrary constants.

Using Lie symmetry analysis, new set of generators of infinitesimal transformations
for Eq. \eqref{v3inF} are obtained and given below:
\begin{align}\label{v3gene}
\xi_X&=\frac{X}{2} (A_3-A_1) +Y A_5 +A_6,\notag\\
\xi_Y&= A_3 Y  +A_4,\notag\\
\xi_Z&=A_1 Z + A_2,\notag\\
\eta_F&= \frac{1}{2} (A_1-A_3)F + \frac{Z}{10}  A_5+A_7,
\end{align}
where $\xi_X, \xi_Y , \xi_Z$ and $\eta_F$ denote generators of infinitesimal
transformations with respect to indicated variable
and $A_1, A_2, A_3, A_4, A_5, A_6$ and $A_7$ are arbitrary constants.

Consequently, this case can be categorized into six subcases discussed below.
\subsubsection{For $A_1 \ne 0, A_2 = 0, A_3= 0, A_4 = 0, A_5 = 0, A_6 = 0$ and $A_7 = 0$ in Eq. \eqref{v3gene} }
We obtain Lagrange's equations as given below
\begin{align}
\frac{dX}{-\frac{X}{2}}=\frac{{dY}}{0}=\frac{{dZ}}{Z}=\frac{{dF}}{\frac{F}{2}}.
\end{align}
The similarity form is 
\begin{align}\label{v3A1Grs}
F=\frac{1}{x}\, G(r, s),\,\,\, \text{with similarity variables}\,\,\, r =X \sqrt{Z}\,\,\,  \text{and}\,\,\, s =Y.
\end{align}
By substituting similarity form into Eq. \eqref{v3inF}, we obtain reduced (1+1)dimensional nonlinear PDE as
\begin{align}\label{v3A1eqinG}
r^4 G_{5r} +&2 (30 G^2 G_r +2 r G_s (1-3 G +3 r G_r )-10 G ( 3 G_r + 6 r G_r^2 +r^2 G_{rrr})\notag \\
+&r (30 G_r^2+ 30 r G_r^3- 2 r G_{rs} + r^2 G_{rrs}+15 G_r (-r G_{rr}+r^2 G_{rrr})))=0.
\end{align}
This equation can be solved numerically.
\subsubsection{For $A_2 \ne 0$, all others are zero in Eq. \eqref{v3gene}}
Corresponding Lagrange's equations as given below
\begin{align}
\frac{dX}{0}=\frac{{dY}}{0}=\frac{{dZ}}{1}=\frac{{dF}}{0}.
\end{align}
The group invariant solution is 
\begin{align}\label{v3A2Grs}
F=G(r, s),
\end{align}
with similarity variables $r =X$ and $s =Y$. By substituting group invariant solution into Eq. \eqref{v3inF}, we have PDE as
\begin{align}\label{v3A2eqinG}
6 G_s G_r+ G_{rrs}=0.
\end{align}
Substituing $G(r,s) = H(\zeta)$ where $\zeta = a\,r +b \,s$ in Eq. \eqref{v3A2eqinG}, we get an ODE in $H$ as
\begin{align}\label{v3A3ode1}
6 H'^2+a H'''=0.
\end{align}
The general solution of \eqref{v3A3ode1} is given as
\begin{align}\label{v3A3H1}
H(\zeta)=a \left(-a^{-1}\right)^{-\frac{1}{3}} WeierstrassZeta \left[\left(-a^{-1}\right)^{\frac{1}{3}} 
(\zeta)+c_6;0,{c_7}\right]+{c_8}.
\end{align}
Hence, using Eqs. \eqref{v3A3H1} and \eqref{v3A3Grs}, we obtain WeierstrassZeta function for Eq. \eqref{kdv} given as 
\begin{align}\label{v3A3u1}
u_4(x,y,z,t)=a \left(-a^{-1}\right)^{-\frac{1}{3}} WeierstrassZeta \left[\left(-a^{-1}\right)^{\frac{1}{3}} 
(a x+b y)+c_6;0,{c_7}\right]+{c_8}.
\end{align}

\begin{figure}[h!]
\centering
\subcaptionbox{}{
\includegraphics[width=0.50\textwidth]{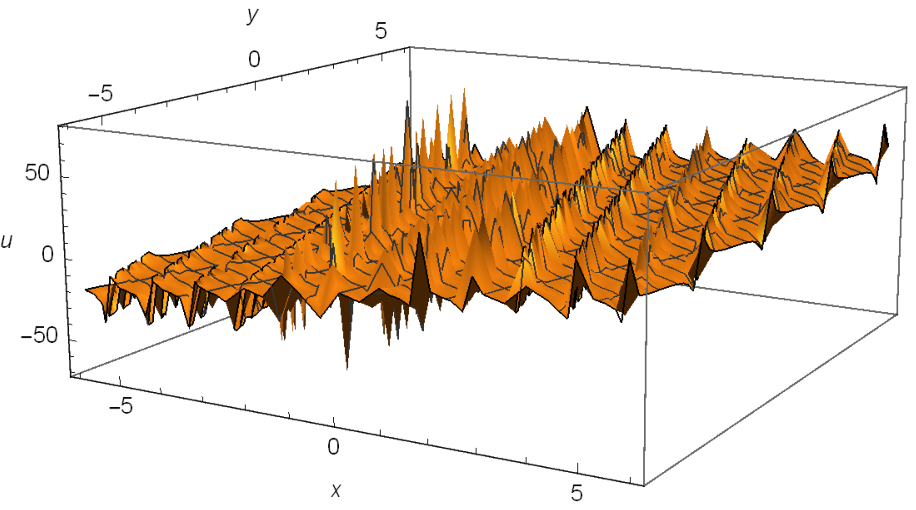}
}%
\hfill
\subcaptionbox{}{
\includegraphics[width=0.30\textwidth]{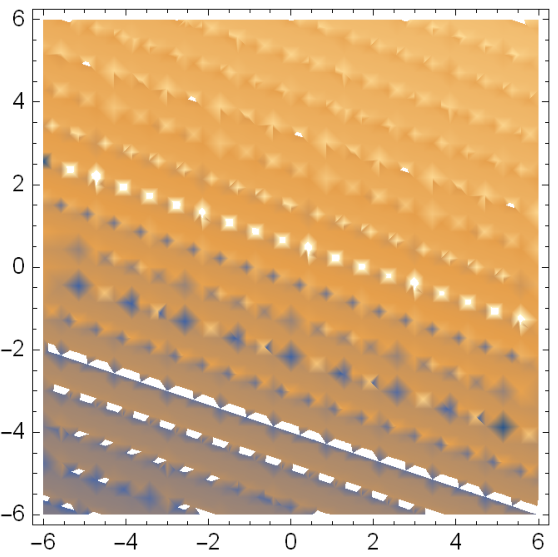}
}%
\hfill 
\caption{Multisoliton solution Eq. \eqref{v3A3u1} of KdV type equation with suitable parameters $a = 3$, $b=1, a_1 = 4, a_2 = 3, a_3 = 9$.  (a) Invariant solution profile shown by WeierstrassZeta function.  (b) The overhead view (density plot) of the solution.}
\label{v3a3}
\end{figure}

Moreover, suppose $H'(\zeta) = f(\zeta)$ in Eq. \eqref{v3A3ode1},  we obtain
\begin{align}\label{v3A3ode2}
6 f(\zeta)^2 + a f''(\zeta) = 0.
\end{align}
The general solution of Eq. \eqref{v3A3ode2} contain WeierstrassP function given as

\begin{align}
f(\zeta)=\left(-a^{-1}\right)^{-a^{-1}}WeierstrassP \left[\left(-a^{-1}\right)^{\frac{1}{3}} \left(\zeta+c_1\right);0,c_2\right].
\end{align}
Hence, WeierstrassP function solution is given as
\begin{align}\label{v3A3u2}
u_5(x,y,z,t) = \left(-a^{-1}\right)^{-\frac{1}{3}}WeierstrassP \left[\left(-a^{-1}\right)^{\frac{1}{3}} \left(a x + b y+c_1\right);0,c_2\right].
\end{align}
\begin{figure}
\centering
\subcaptionbox{}{
\includegraphics[width=0.50\textwidth]{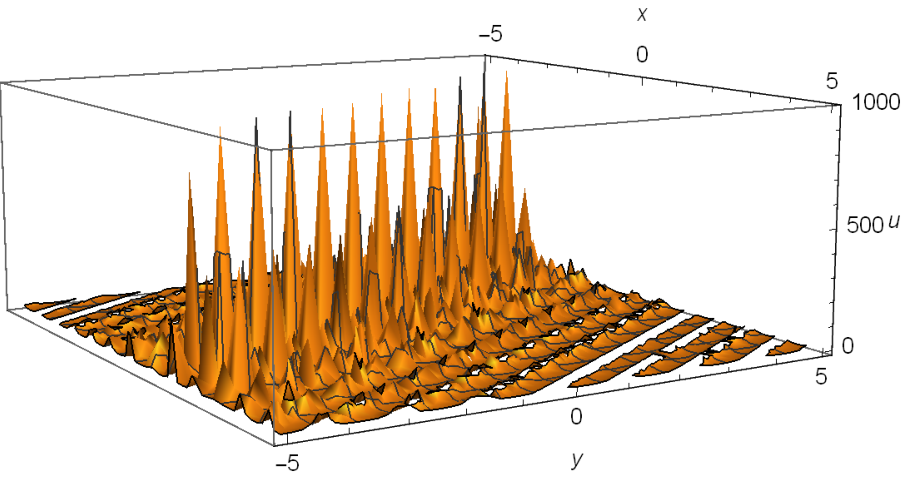}
}%
\hfill
\subcaptionbox{}{
\includegraphics[width=0.30\textwidth]{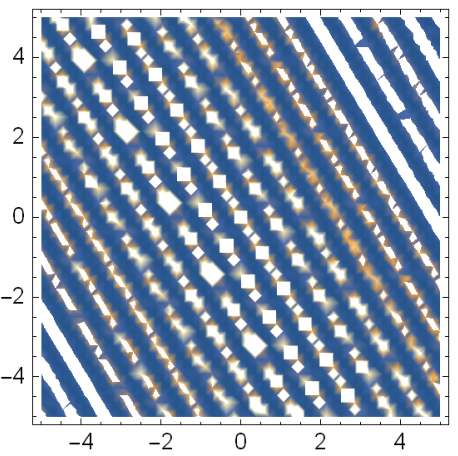}
}%
\hfill 
\caption{Multisoliton solution Eq. \eqref{v3A3u2} of KdV type equation with suitable parameters $a = 5, b=3, c_1=4, c_2 = 5$. (a) Invariant solution profile for WeierstrassP function.  (b) The overhead view (density plot) of the solution.}
\label{v3a4a}
\end{figure}
Now integrating Eq. \eqref{v3A3ode2}, obtain a new ordinary differential equation given as
\begin{align}\label{v3A3ode3}
2f^3 +\frac{1}{2} a f'^2 = \alpha.
\end{align}
When we take $\alpha=0$ and $a = -k^2$ two general solutions arises, they are given as
\begin{align}
f(\zeta) = \frac{4 k^2}{\left(c_1 k \pm 2 \zeta \right){}^2}
\end{align}
Using back substitution, we obtain rational function solutions for Eq. \eqref{kdv} are given by
\begin{align}
u_{6,7}(x,y,z,t)& = \frac{2 k^2}{\alpha_1 k\pm2 (a x +b y)}.
\end{align}
Also, for $a = k^2$ two general solutions arises, they are given as
\begin{align}
f(\zeta) = \frac{4 k^2}{\left(c_1 k \pm 2 i\zeta \right){}^2}.
\end{align}
corresponding rational function solutions
\begin{align}
u_{8,9}(x,y,z,t)& = \frac{2 k^2}{\alpha_1 k\pm  2 i (a x +b y)}.
\end{align}

\subsubsection{For $A_3 \ne 0$, all others are zero in Eq. \eqref{v3gene}}
In this case, we get associated Lagrange's system as follows
\begin{align}\label{v3A3lag}
\frac{dX}{\frac{X}{2}}=\frac{{dY}}{Y}=\frac{{dZ}}{0}=\frac{{dF}}{-\frac{F}{2}}.
\end{align}
The group invariant solution is 
\begin{align}\label{v3A3F}
F=\frac{1}{\sqrt{Y}} G(r, s),
\end{align}
with similarity variables $r =\frac{X}{\sqrt{Y}}$ and $s =Z$. By substituting group invariant solution into Eq. \eqref{v3inF}, we have (1+1) dimensional nonlinear PDE as
\begin{align}\label{v3A3Grs}
2G_{rrrrs}-3G_{rr}-2G_r (3G+3(r-20G_s)G_r-20G_{rrs})-(r-20G_s)G_{rrr}=0.
\end{align}
Lie group analysis method gives the following generators
of infinitesimal transformations when it is applied
on Eq. \eqref{v3A3Grs}.
\begin{align}\label{v3a2}
\xi_r =-\frac{r}{2} b_1,\,\,
\xi_s =b_1 s+b_2,\,\,
\xi_G =\frac{G}{2}b_1,
\end{align}
where $b_1$ and $b_2$ are arbitrary constants.
The similarity solution $G(r, s)$ can be written in the
following similarity form
\begin{align}\label{eq64}
G(r,s)=\sqrt{s}\, R(w),
\end{align}
with similarity variable $w= r \sqrt{s}$. 
This transformation
reduces Eq. \eqref{v3A3Grs} into following ODE
\begin{align}\label{v3A3inw}
wR^{(5)}+5R^{(4)}-3R''+6R' [(R+wR')(10R'-1)+10R'']+10R+w(30R'-1)]R'''=0.
\end{align}
Eq. \eqref{v3A3inw} is a complicated nonlinear ODE and cannot be solved in general. Anyhow, assuming the adequate values of arbitrary constants, some particular solutions are given below 
\begin{align}\label{eq66}
R(w) = \alpha, \,\,\,\,
R(w) = \frac{\beta}{w},\,\,\,\,
R(w) = \frac{1}{10}w+\gamma,
\end{align}
where $\alpha, \beta$ and $\gamma$ are arbitrary constants.
Ultimately, using Eqs. \eqref{eq66}, \eqref{eq64},  \eqref{v3A3F} 
in Eq. \eqref{v3uF} corresponding $u$ can be given as
\begin{align}
u_{10}(x,y,z,t) &= \alpha \sqrt{\frac{z}{y}},\\
u_{11}(x,y,z,t) &=  \frac{\beta}{x},\\
u_{12}(x,y,z,t) &= \frac{x z}{10 y}+\gamma \sqrt{\frac{z}{y}}.
\end{align}

\subsubsection{For $A_4 \ne 0$, all others are zero in Eq. \eqref{v3gene}}
In this case, we obtain characteristic equations as follows
\begin{align}
\frac{dX}{0}=\frac{{dY}}{1}=\frac{{dZ}}{0}=\frac{{dF}}{0}.
\end{align}
The group invariant solution is 
\begin{align}\label{v3A4Grs1}
F=G(r, s),
\end{align}
with similarity variables $r =X$ and $s =Z$. 
By substituting group invariant solution into Eq. \eqref{v3inF}, we have
(1+1) dimensional nonlinear PDE as
\begin{align}\label{v3A4Grs}
20G_r G_{rrs} +10G_s (6G_r^2+G_{rrr})+G_{rrrrs}=0.
\end{align}

Lie group analysis method gives the following generators
of infinitesimal transformations when it is applied
on Eq. \eqref{v3A3Grs}.
\begin{align}\label{v3a4}
\xi_r =b_1 r+b_2,\,\,
\xi_s =f(s), \,\,
\xi_G =-G b_1+b_3,
\end{align}
where $b_1, b_2$ and $b_3$ are arbitrary constants.
The similarity solution $G(r, s)$ can be written in the
following similarity form
\begin{align}\label{eq74}
G(r,s)=\frac{b_3 r+R(w)}{b_1 r + b_2},\,\,\,\,\,\, \text{with similarity variable}\,\,\,\, 
w= \frac{b_1 r+b_2}{b_1 s}.
\end{align}
This transformation reduces Eq. \eqref{v3A4Grs} into following ODE
\begin{align}\label{v3A4inw}
b_1^4 w^4 R^{(5)}+&20 b_1^2 w^2 R^{(3)} \left( b_1 w R'-b_1 R+b_2 b_3\right)+10 R' b_1^3 w^3 R^{(3)}-30 b_1^3 w^2 R' R''\notag \\
-&60 R' \left( b_1 R-b_1 w R' \right) \left(b_1^2+b_1 w R' - b_1 R + 2 b_2 b_3 \right)+60  b_2 b_3 (b_1^2+b_2 b_3)R'=0.
\end{align}
Eq. \eqref{v3A4inw} is a complicated nonlinear ordinary differential equation and cannot be solved in general. Anyhow assuming the adequate values of arbitrary constants, some particular solutions are given below
\begin{align}\label{eq76}
R(w) = \alpha, \,\,\,\,
R(w) = \beta w + \frac{b_2 b_3}{b_1},\,\,\,\,
R(w) = \gamma w+\frac{b_2 b_3}{b_1}+b_1,
\end{align}
where $\alpha, \beta$ and $\gamma$ are arbitrary constants.
Using Eqs. \eqref{eq76}, \eqref{eq74}, \eqref{v3A4Grs1} in Eq. \eqref{v3uF}, corresponding rational function solution $u$ for (3+1) kdV type equation can be given as
\begin{align}
u_{13}(x,y,z,t) &= \frac{\alpha +b_3 x}{b_1 x+b_2},\\
u_{14}(x,y,z,t) &= \frac{b_3 z+\beta}{b_1 z},\\
u_{15}(x,y,z,t) &= \frac{\gamma}{b_1 z} +\frac{b_1}{b_1 x+b_2} + \frac{b_3}{b_1}.
\end{align}

\subsubsection{For $A_5 \ne 0$, all others are zero in Eq. \eqref{v3gene}}
We write Lagrange's equations as given below
\begin{align}
\frac{dX}{Y}=\frac{{dY}}{0}=\frac{{dZ}}{0}=\frac{{dF}}{\frac{Z}{10}}.
\end{align}
The group invariant solution is 
\begin{align}\label{v3A5Grs}
F=\frac{X Z}{10 Y} +G(r, s),
\end{align}
with similarity variables $r =Y $ and $s =Z$. By substituting group invariant solution into Eq. \eqref{v3inF}, we have PDE as
\begin{align}\label{v3A5eqinG}
s G_s +r G_r =0. 
\end{align}
The general solution is given as 
\begin{align}\label{eq83}
G(r,s) = f\left( \frac{s}{r}\right).
\end{align}
Hence, using Eqs. \eqref{eq83}, \eqref{v3A5Grs}, in Eq. \eqref{v3uF} invariant solution $u$ with $f$ as arbitrary function is given as
\begin{align}
u_{16}(x,y,z,t) = \frac{x z}{10 y} + f\left( \frac{z}{y}\right).
\end{align}
\subsubsection{For $A_6 \ne 0\,\,\, \text{and}\,\,\,  A_7 \ne 0$ in Eq. \eqref{v3gene} }
In both cases, nontrivial solutions for Eq. \eqref{kdv} does not exist.
\subsection{\noindent \textbf{\textit{Vector field $v_4$:}}}
The associated Lagrange system for
\begin{align}
v_4= \frac{3t}{10} \frac{\partial}{\partial y}
+y \frac{\partial}{\partial z}
+\frac{x}{20} \frac{\partial}{\partial u},  \label{eqv4}
\end{align}
is given by
\begin{align}\label{v4lag}
\frac{dx}{0}=\frac{dy}{\frac{3t}{10}}=\frac{dz}{y}=\frac{dt}{0}=\frac{du}{\frac{x}{20}}.
\end{align}
Using Eq. \eqref{v4lag}, the similarity form is given by
\begin{align*}
u=\frac{x y}{6t}+ F(X,Y,T)\,\,\, \text{ where similarity variables are}\,\,\,X=x, Y= z-\frac{5 y^2}{3t} , T=t.
\end{align*}
Substituting the similarity solution in Eq. \eqref{kdv}, we obtain the following reduction equation 
\begin{align}\label{v4inF}
X F_X+ T (F_T+ 20 F_X F_{XXY}+10F_Y (6F_X^2+F{XXX})+F_{XXXXY})=0.
\end{align}
New set of generaotors for Eq. \eqref{v4inF} are
\begin{align}\label{v4gene}
\xi_X = \frac{X}{4}(A_1-A_2)+A_4 T,\,\, \xi_Y =A_2 T+A_3, \,\,\,\, \xi_T =A_1 T,\,\,\,\, \eta_F = \frac{F}{4}(A_2-A_1) +A_5,
\end{align}
where $A_1, A_2, A_3, A_4$, and $A_5$ are arbitrary constants.

\begin{table}
\begin{center}
\caption{Reduced equations and Invariant solutions of the (3+1)-KdV type equation for case $v_4$.}
\begin{tabular}{ cccc } 
\hline \hline
Subcase 			& Similarity variables 		& Reduced equations & Invariant solutions \\ 
\hline\hline
$A_1 \ne 0$, and 	& $r=X T^{-\frac{1}{4}}, s = Y$	& $ G + G_r(-3r-80 G_{rrs}) $ &  $u=\frac{x y}{6 t}-\frac{x \sqrt{3 b_1 t z-5 b_1 y^2+3 b_2 t}}{6 \sqrt{5} \sqrt{-b_1} t}$\\
other $A_i$'s are zero  				& $F =T^{-\frac{1}{4}}G(r,s) $ 	& $-40G_s(6G_r^2+G_{rrr})-4G_{rrrrs}=0$ & \\ \\
$A_2 \ne 0$, and	& $r = X Y^{\frac{1}{4}}, s = T$	& $4 r G_r+s [ 4 G_s+5(2G (6 G_r^2+G_{rrr})+6G_r(2G_{rr}$&$u=\frac{x y}{6 t}-\frac{x \sqrt{5 \left(5 y^2-3 t z\right)}}{30 t}$  \\
 other $A_i$'s are zero		& $F= Y^{-\frac{1}{4}}G(r,s)$ 	&$+r(2G_r^2+G_{rrr}))+G_{rrrr})+rG_{rrrrr}]=0$&\\ \\
 				
  $A_3 \ne 0$, and	& $r = X, s = T$ 		&$sG_s+rG_r=0$  & $ u=\frac{x y}{6t} + f\left( \frac{x}{t}\right)$ \\
other $A_i$'s are zero& $F = G(r,s)$ 			  				& &\\ \\
 				
  $A_4 \ne 0$, and	& $r = Y, s = T$ 		&  $G_s=0$ &$u=\frac{x y}{6t} + f\left(z- \frac{5y^2}{3t}\right)$ \\
other $A_i$'s are zero& $F= G(r,s)$ 			  				& & \\ \\
 				
 $A_5 \ne 0$, and	&   & No Solution  & No solution \\
 other $A_i$'s are zero&   &   &  \\
  \hline
\end{tabular}
\label{tabv4}
\end{center}
\end{table}

Moreover, reduced equations and invariant solutions for subcases are furnished by Table \ref{tabv4}.
\subsection{\noindent \textbf{\textit{Vector field $v_5$:}}}
The associated Lagrange system for
\begin{align}
v_5= \frac{-x}{4}\frac{\partial}{\partial x}
+\frac{y}{2} \frac{\partial}{\partial y}
+z \frac{\partial}{\partial z}
+\frac{u}{4} \frac{\partial}{\partial u},  \label{eqv5}
\end{align}
is given by
\begin{align}
\frac{dx}{\frac{-x}{4}}=\frac{{dy}}{\frac{y}{2}}=\frac{{dz}}{z}=\frac{{dt}}{0}=\frac{{du}}{\frac{u}{4}}.
\end{align}
The similarity variables are $X=x z^{\frac{1}{4}}, Y=\frac{y}{\sqrt{z}} , T=t$, and the group-invariant solution is $u=z^{\frac{1}{4}} F(X,Y,T)	$.  Substituting the group-invariant solution in Eq. \eqref{kdv}, we obtain the following reduction equation 
{\small
\begin{align}\label{v5inF}
XF_{XXXXX}&-2YF_{XXXXY}+5F_{XXXX}+4 F_T+60 F F_X^2+60 X F_X^3 +60 F_X F_{XX}+4 F_{XXY} \notag\\
&-40Y F_X F_{XXY}+10F F_{XXX}+30X F_X F_{XXX}+4 F_Y (6F_X(1-5YF_X)-5YF_{XXX})=0.
\end{align}}
The equation can be solved numerically.
\subsection{\noindent \textbf{\textit{Vector field $v_6$:}}} 
The associated Lagrange system for 
\begin{align}
v_6= \frac{\partial}{\partial z}, \label{eqv6}
\end{align}
is given by
\begin{align}
\frac{dx}{0}=\frac{{dy}}{0}=\frac{{dz}}{1}=\frac{{dt}}{0}=\frac{{du}}{0}.
\end{align}
The similarity variables are $X=x, Y=y, T=t$, and the group-invariant solution is $u= F(X,Y,T)	$.  Substituting the group-invariant solution in Eq. \eqref{kdv}, we obtain the following reduction equation 
\begin{align}\label{v6inF}
F_{XXY}+6F_X F_Y+F_T=0.
\end{align}
The general solution of Eq. \eqref{v6inF} is obtained and given as
\begin{align}\label{v6eq72}
F(X,Y,T) = c_1 \tanh \left(c_1 X-\frac{c_3 Y}{4 c_1^2}+c_3 T+c_4\right)+c_5. 
\end{align}
Back substitution of Eq. \eqref{v6eq72}, we obtain kink wave solution of (3 + 1)-KdV type equation given as
\begin{align}\label{v6eq72a}
u_{17}(x,y,z,t)=c_1 \tanh \left(c_1 x-\frac{c_3 y}{4 c_1^2}+c_3 t+c_4\right)+c_5.
\end{align}
New set of generaotors for Eq. \eqref{v6inF} are given below:
\begin{align}
\xi_X &= \frac{X}{2}(A_1-A_4)+A_6 T+A_7,\,\,\,\,
\xi_Y =A_3 T+A_4 Y+A_5, \notag\\
\xi_T &=A_1 T+A_2, \,\,\,\,\,\,\,\,\,\,\,\,\,\,\,\,\,\,\,\,\,\,\,\,\,\,\,\,\,\,\,\,\,\,\,\,\,\,\,\,\,\,\,\,
\eta_F = \frac{F}{2} (A_4-A_1)+\frac{X}{6} A_3+\frac{Y}{6}A_6 +A_8.
\end{align}
Consequently, reduced equations for various subcases are provided by Table \ref{tabv6}.

\begin{table}[h!]
\begin{center}
\caption{Reduced equations and invariant solutions of the (3+1)-KdV type equation for case $v_6$.}
\begin{tabular}{ cccc } 
 \hline  \hline
  Subcase 			& Similarity variables 			  		& Reduced equations & Invariant solutions  \\ 
  \hline\hline
$A_1 \ne 0$, and 	& $r=X T^{-\frac{1}{2}}, s = Y$	& $ G + G_r(r-12 G_{s})-2 G_{rrs}=0$ & $u= \frac{x }{6t}(y+b)$ \\
other $A_i$'s are zero& $F =T^{-\frac{1}{2}}G(r,s)$ 			  				& & \\ \\
$A_2 \ne 0$, and	& $r = X, s = Y$	&   $6G_sG_r+G_{rrs}=0$, & $u_5$ and $u_4$ \\
other $A_i$'s are zero& $F= G(r,s)$ 			  				& & \\ \\
 				
$A_3 \ne 0$, and	& $r = X, s = T$ 		&$sG_s+rG_r=0$ & $u=\frac{xy}{6t}+f\left( \frac{t}{x} \right)$  \\ 
other $A_i$'s are zero& $F=\frac{xy}{6t}+ G(r,s)$ 			  				& & \\ \\
 				
$A_4 \ne 0$, and	& $r =X\sqrt{Y}, s = T$ 		&$2G_s+6G_r(G+rG_r)$ & $u=\frac{\alpha \sqrt{y}}{\sqrt{B_1 t+B_2}}+\frac{B_1 x y}{6 (B_1 t+B_2)}$   \\ 
other $A_i$'s are zero& $F=\sqrt{Y} G(r,s)$ 			  				&$+3G_{rr}+rG_{rrr}=0$ & \\ \\
 				
$A_5 \ne 0$, and	& $r = X, s = T$ 		&  $G_s=0$ &$u=f\left(x \right)$\\ 
other $A_i$'s are zero& $F= G(r,s)$ 			  				& & \\ \\

$A_6 \ne 0$, and	& $r = Y, s = T$ 		&$sG_s+rG_r=0$ &$u= \frac{xy}{6t}+ f(\frac{t}{y})$   \\ 
 other $A_i$'s are zero& $F=\frac{xy}{6t}+ G(r,s)$ 			  				& & \\ \\

$A_7 \ne 0$, and	& $r = Y, s = T$ 		&  $G_s=0$ &$u= f\left(y\right)$\\ 
other $A_i$'s are zero& $F= G(r,s)$ 			  				& & \\ \\
 
$A_8 \ne 0$, and	&  		&		No solution	   & \\ 
other $A_i$'s are zero &  		&		   & \\ 
\hline
\end{tabular}
\label{tabv6}
\end{center}
\end{table}

\subsection{\noindent \textbf{\textit{Vector Field $v_7$:}}}
For the infinitesimal generator
\begin{eqnarray*}
v_7= \frac{\partial}{\partial y},
\end{eqnarray*}
we have following characteristic equation
\begin{align}\label{v7gen}
\frac{dx}{0}=\frac{{dy}}{1}=\frac{{dz}}{0}=\frac{{dt}}{0}=\frac{{du}}{0}.
\end{align}
The group invariant solution is
\begin{align}\label{eq99}
u= F(X,Z,T),
\end{align}
where $X=x, Z=z, T=t$.
Substituting the group-invariant solution in Eq. \eqref{kdv}, we obtain the following reduction equation
\begin{align}\label{v7inF}
F_{XXXXZ}+20 F_{XXZ} F_X+ 10  F_Z (6 F_X^2+ F_{XXX})  +F_T=0.
\end{align}
The general solution of Eq. \eqref{v7inF} are given as
\begin{align}\label{eq101}
F(X,Z,T)=c_5+c_1 \tanh \left(c_1 X-\frac{c_3}{16 c_1^4 }Z+c_3 T+c_4\right).
\end{align}
Hence, substituting the value of $F$ in Eq. \eqref{eq99} we obtain kink wave solution of our original Eq. \eqref{kdv}, 
\begin{align}
u_{18}(x,y,z,t)&=c_5+c_1 \tanh \left(c_1 x-\frac{c_3}{16 c_1^4 }z+c_3 t+c_4\right).
\end{align}
New set of generaotors for Eq. \eqref{v7inF} are
\begin{align}\label{v7gene}
\xi_X = \frac{X}{4}(A_1-A_3)+A_5,\,\,\,\, \xi_Z = A_3 Z+A_4, \,\,\,\, \xi_T =A_1 T+A_2,\,\,\,\, \eta_F = \frac{F}{4}(A_3-A_1) +A_6,
\end{align}
where $A_1, A_2, A_3, A_4, A_5$ and $A_6$ are arbitrary constants.
Consequently, this case also be categorized into the following subcases:
\subsubsection{For $A_1=0, A_3=0$ in Eq. \eqref{v7gene}}
In this subcase, characteristic equations are given by
\begin{align}
\frac{dX}{A_5}=\frac{{dZ}}{A_4}=\frac{{dT}}{A_2}=\frac{dF}{A_6}.
\end{align}
The group invariant solution is 
\begin{align}
F=\left( \frac{A_6}{A_2}\right) T+G(r,s),\,\,\, \text{where}\,\,\, r=X-\left( \frac{A_5}{A_2}\right) T, s=Z-\left( \frac{A_4}{A_2}\right) T.
\end{align}
 
Then, by substituting group invariant solution in Eq. \eqref{v7inF}, we have following reduced equation which is given as follows
\begin{align}\label{v7Grs}
A_6-G_r(A_5-20 A_2 G_{rrs})+G_s(-A_4+10 A_2 (6G_r^2+G_{RRR})+A_2 G_{rrrrs}=0. 
\end{align}
Again, new set of generators are 
\begin{align}
\xi_r = b_2,\,\,\,\, \xi_s = b_1, \,\,\,\, \eta_G = b_3,
\end{align}
where $b_1, b_2$ and $b_3$ are integral constants
and group invariant solution is
\begin{align}
G=\left( \frac{b_3}{b_1}\right) s+R(w),\,\,\,\, \text{where} \,\,\,w=r-\left( \frac{b_2}{b_1}\right) s.
\end{align}

Substituting in Eq. \eqref{v7Grs}, we obtained ordinary differential equation as
\begin{align}\label{v7inR}
A_2 b_2 R^{(5)}+(10 A_2 (R^{(3)}+6 R'^2)-A_4) (b_3-b_2 R' )-R' (20 A_2 b_2 R^{(3)}+A_5 b_1 )+A_6 b_1=0.
\end{align} 
Unfortunately, we could not find general solution of Eq. \eqref{v7inR}, but we manage to find one particular solution given as follows
\begin{align}
R(w) = b\, w + c,
\end{align} 
where $b$ and $c$ are constant of integration. Here, $b$ in terms of other constants given as follows
{\small
\begin{align}
b=-&\frac{\sqrt[3]{5A_8+\frac{1}{216} \sqrt{A_7}}}{30 A_2 b_2 }+\frac{5^{2/3} \left(-20 A_2 b_3^2-A_4 b_2^2+A_5 b_1 b_2\right)}{30 b_2 \sqrt[3]{A_8+\frac{\sqrt{A_7}}{1080}}}+\frac{b_3}{3 b_2},\notag \\
\text{where} \notag\\
A_7=&29160000 A_2^4 \left(40 A_2 b_3^3-3 b_2 b_3 (2 A_4 b_2+A_5 b_1)+9 A_6 b_1 b_2^2\right)^2+\left(-3600 A_2^2 b_3^2-180 A_2 b_2 (A_4 b_2-A_5 b_1)\right)^3, \notag \\
A_8 =& -200 A_2^3 b_3^3+15 A_2^2 b_2 (2 A_4 b_2 b_3+A_5 b_1 b_3-3 A_6 b_1 b_2). \notag
\end{align}
}
Hence, rational function solution for Eq. \eqref{kdv} is given as
\begin{align}
u_{19}(x,y,z,t)=\frac{A_2 (b_1 (b x+c)+z (b_3-b b_2))+t (A_4 b b_2-A_4 b_3-A_5 b b_1+A_6 b_1)}{A_2 b_1}.
\end{align}

\subsection{\noindent \textbf{\textit{Vector field $v_8$:}}}
For the infinitesimal generator
\begin{eqnarray}
v_8 = t\frac{\partial}{\partial x}+ \frac{y}{6}\frac{\partial}{\partial u}, 
\end{eqnarray}
we have following characteristic equation
\begin{align}
\frac{dx}{t}=\frac{{dy}}{0}=\frac{{dz}}{0}=\frac{{dt}}{0}=\frac{{du}}{\frac{y}{6}}.
\end{align}
We reduce Eq. \eqref{kdv} into following invariant form with new invariant variables
\begin{align}
u(x,y,z,t) = \frac{x y}{6 t} + F(Y,Z,T),
\end{align}
where $Y=y, Z=z, T=t$.

\begin{figure}
\centering
\subcaptionbox{$z = 0.9654, t=6.$}{\includegraphics[width=0.300\textwidth]{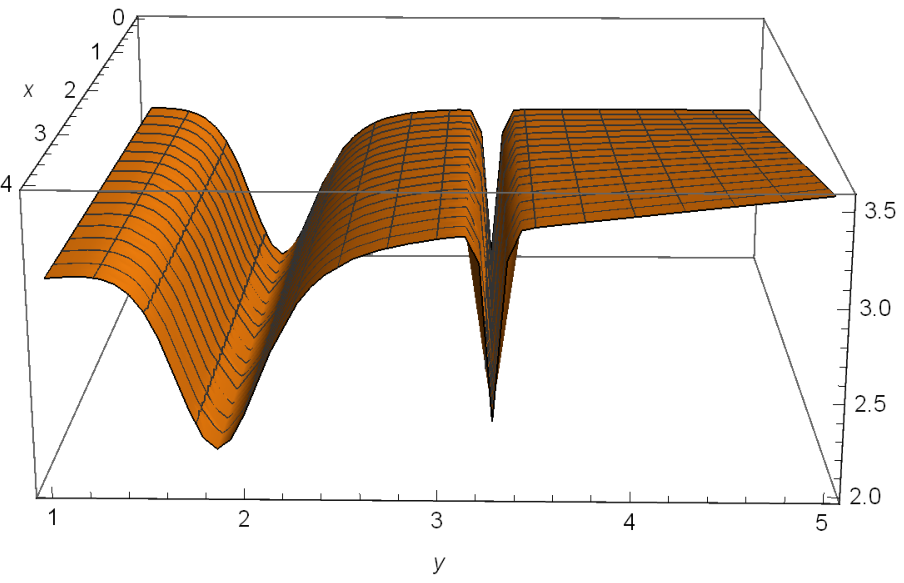}}%
\hfill
\subcaptionbox{$x = 3.5, z = 0.9654$.}{\includegraphics[width=0.30\textwidth]{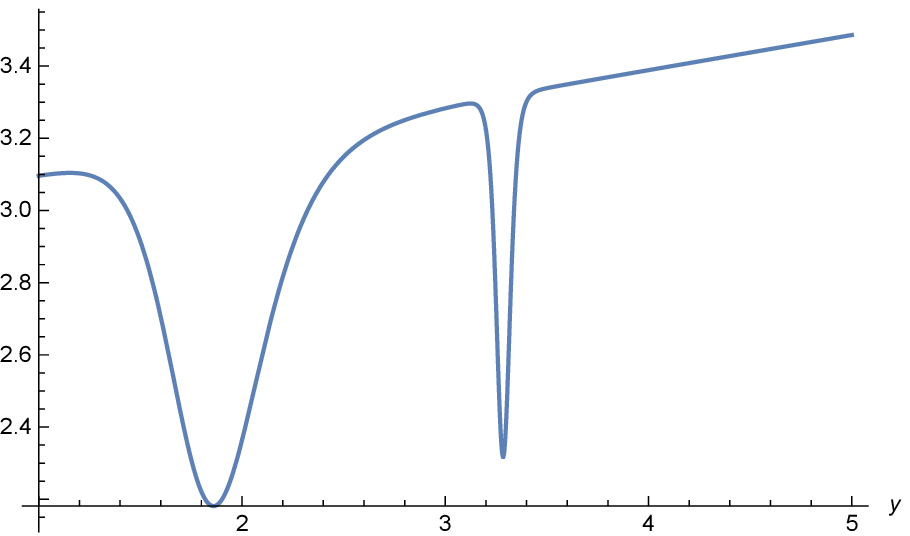}}%
\hfill 
\subcaptionbox{$z = 0.96541, t = 6$.}{\includegraphics[width=0.20\textwidth, angle =-90]{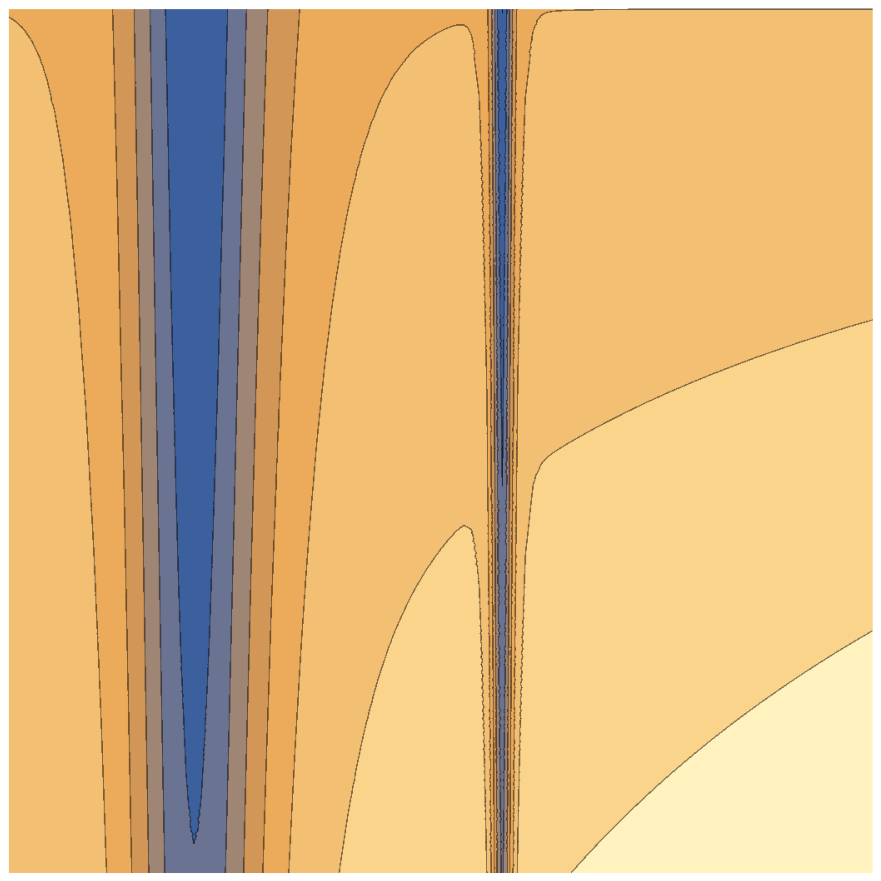}}%
\hfill 
\caption{Solitary waves solution profiles of Eq. \eqref{uwithTanh}.
(a) Doubly soliton. 
(b) The wave propagation pattern of the wave along the $x$-axis.  
(c) Correspoding coutour plot.}
\label{f1v8a}
\end{figure}
Substituting the group invariant solution in Eq. \eqref{kdv}, we obtain the following reduction equation
\begin{align}\label{v8inF}
3 T^2 F_T +5Y^2 F_Z+3 Y T F_Y =0 .
\end{align}
The general solution of Eq. \eqref{v8inF} are given as
\begin{align}
F(Y,Z,T)= g\left(\frac{T}{Y},\frac{3ZT-5Y^2}{3T}\right).
\end{align}
Hence, by back substitution, we shall get invariant solution of our original Eq. \eqref{kdv} as 
\begin{align}\label{u15}
u_{20}(x,y,z,t)&=\frac{xy}{6t}+g\left(\frac{t}{y},\frac{3zt-5y^2}{3t}\right).
\end{align}
Travelling wave solutions of Eq. \eqref{u15} are given below 
{\small
\begin{align}
u_{20a}(x,y,z,t) = \frac{x y}{6 t}&+\text{tanh}^2\left(\frac{3 t z-5 y^2}{3 y}\right)+ \tanh^2\left(\frac{2 t^3}{y^3}+\frac{2 \left(3 t z-5 y^2\right)}{t}\right)+\text{tanh}^2\left(\frac{\left(3 t z-5 y^2\right)^2}{9 t^2}+\frac{2 t^2}{y^2}+1\right),\label{uwithTanh}
\end{align}
\begin{align}
u_{20b}(x,y,z,t) = \frac{x y}{6 t}&+\text{sech}^2\left(\frac{3 t z-5 y^2}{3 y}\right)+ \text{sech}^2\left(\frac{2 t^3}{y^3}+\frac{2 \left(3 t z-5 y^2\right)}{t}\right)+\text{sech}^2\left(\frac{\left(3 t z-5 y^2\right)^2}{9 t^2}+\frac{2 t^2}{y^2}+1\right).\label{uwithSech}
\end{align}
}
\begin{figure}
\centering
\subcaptionbox{$z = 0.96541, t=6$.}{\includegraphics[width=0.300\textwidth]{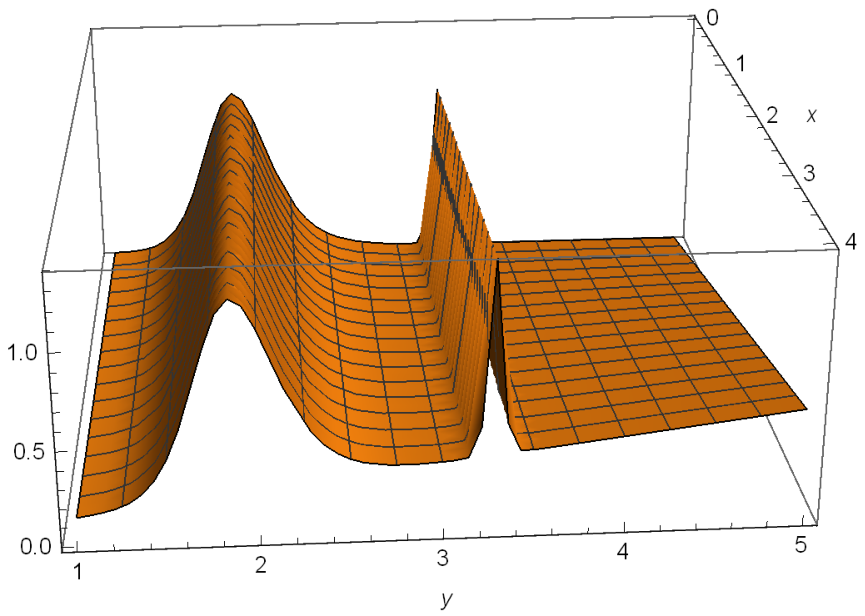}}%
\hfill
\subcaptionbox{$x = 3.5, z = 0.9654$.}{\includegraphics[width=0.35\textwidth]{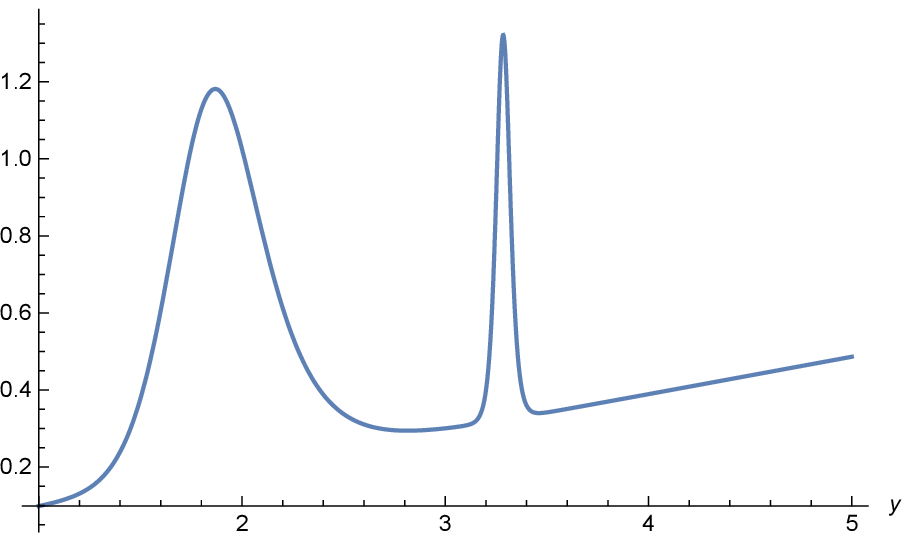}}%
\hfill 
\subcaptionbox{$z = 0.96541, t = 6$.}{\includegraphics[width=0.20\textwidth, angle =-90]{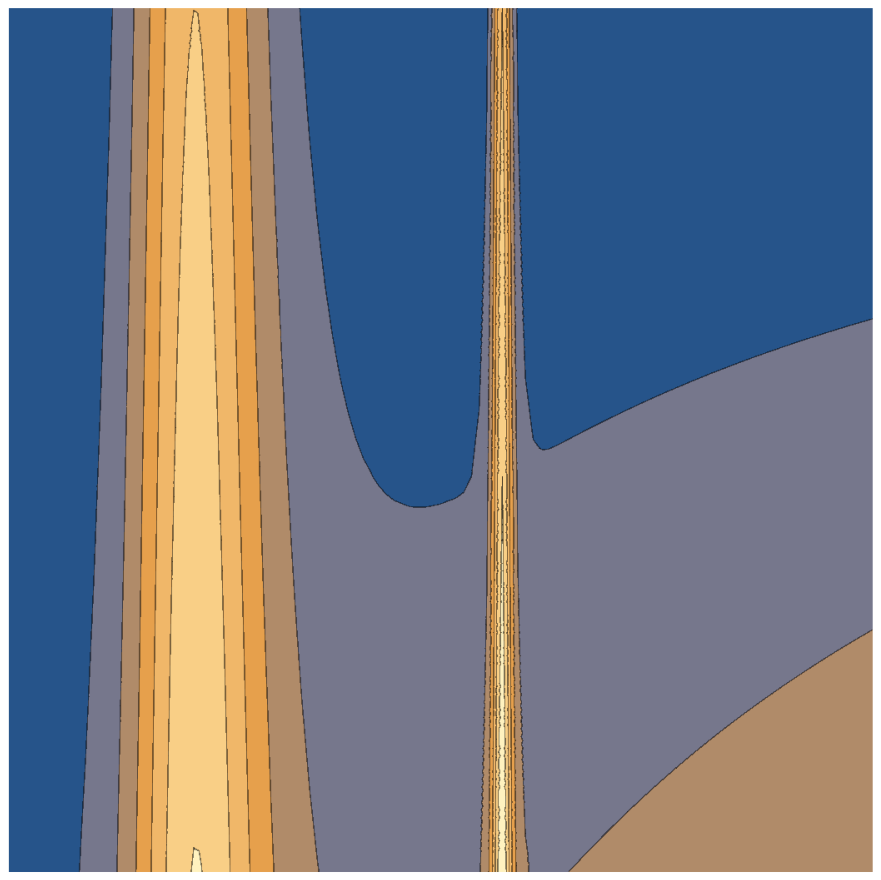}
}%
\hfill 
\caption{Solitary waves solution profiles of Eq. \eqref{uwithSech}.
(a) Doubly soliton. 
(b) The wave propagation pattern of the wave along the $x$-axis. 
(c) Corresponding contour plot.}
\label{f1v6A4b}
\end{figure}

\subsection{\noindent \textbf{\textit{Vector field $v_9$:}}}
For the infinitesimal generator
\begin{eqnarray}
v_9 =  y\frac{\partial}{\partial x}+\frac{z}{10}\frac{\partial}{\partial u}, 
\end{eqnarray}
we have following characteristic equation
\begin{align}
\frac{dx}{y}=\frac{{dy}}{0}=\frac{{dz}}{0}=\frac{{dt}}{0}=\frac{{du}}{\frac{z}{10}}.
\end{align}
We reduce Eq. \eqref{kdv} into following invariant form with new invariant variables
\begin{align}
u(x,y,z,t) = \frac{x z}{10 y} + F(Y,Z,T),
\end{align}
where $Y=y, Z=z, T=t$.
Substituting the group-invariant solution in Eq. \eqref{kdv}, we obtain the following reduction equation
\begin{align}\label{v9inF}
 F_T + \frac{3Z}{5Y^2} (ZF_Z+YF_Y)=0.
\end{align}
The general solution of Eq. \eqref{v8inF} are given as
\begin{align}
F(Y,Z,T)= g\left(\frac{Z}{Y},\frac{3ZT-5Y^2}{3Z}\right).
\end{align}
Hence, by back substitution we shall get invariant solution of our main equation 
\begin{align}\label{u16}
u_{21}(x,y,z,t)&=\frac{xz}{10y}+g\left(\frac{z}{y},\frac{3zt-5y^2}{3z}\right).
\end{align}
\begin{figure}[h!]
\centering
\subcaptionbox{$z = 1,t=15$}{\includegraphics[width=0.300\textwidth]{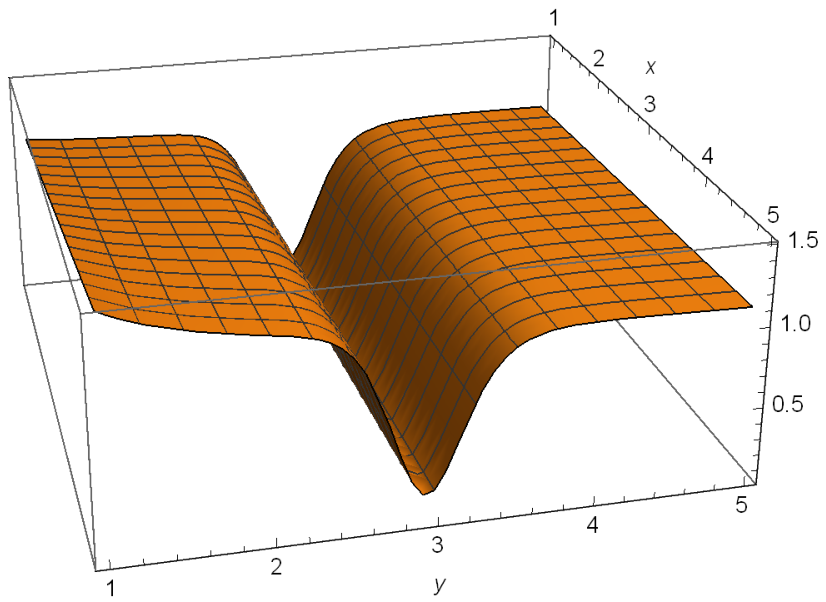}}%
\hfill
\subcaptionbox{$x=2, z = 1, t = 14$}{\includegraphics[width=0.35\textwidth]{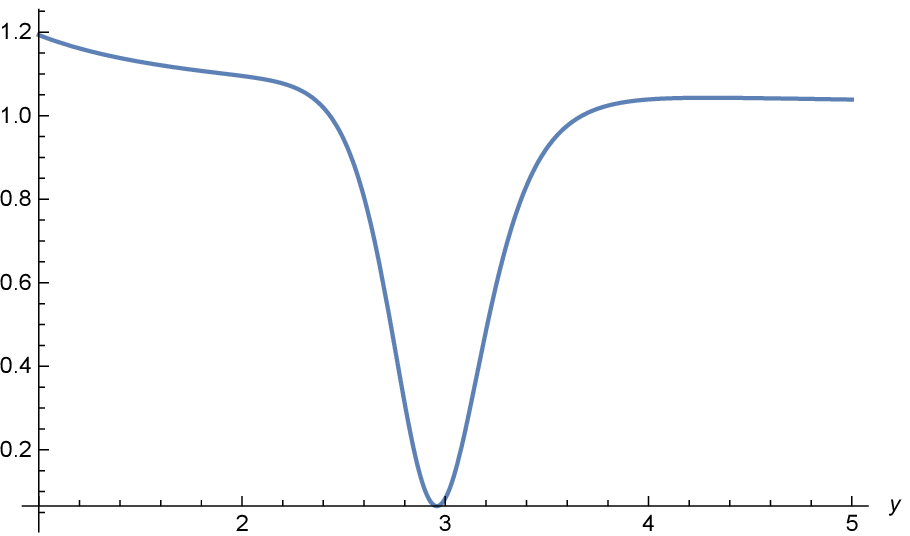}}%
\hfill 
\subcaptionbox{$z = 1, t = 15$}{\includegraphics[width=0.20\textwidth, angle=180]{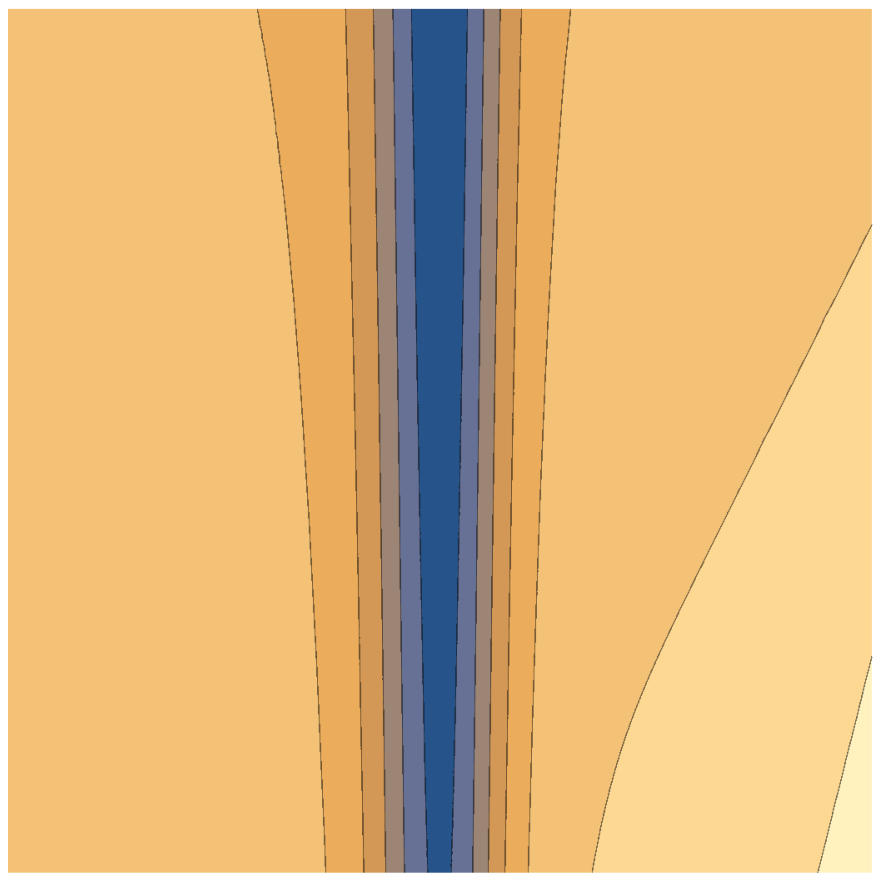}}%
\hfill 
\caption{Solitary waves solution profiles of Eq. \eqref{u16} with $f=\tanh \left[\frac{3 t z-5 y^2}{3y}\right]^2$.
(a) Perspective view of the real part of the single soliton solution. 
(b) The wave propagation pattern of the wave along the $y$-axis. 
(c) The corresponding contour plot. 
}
\label{f1v9a}
\end{figure}

\begin{figure}
\centering
\subcaptionbox{$z = 1, t = 14$}{\includegraphics[width=0.300\textwidth]{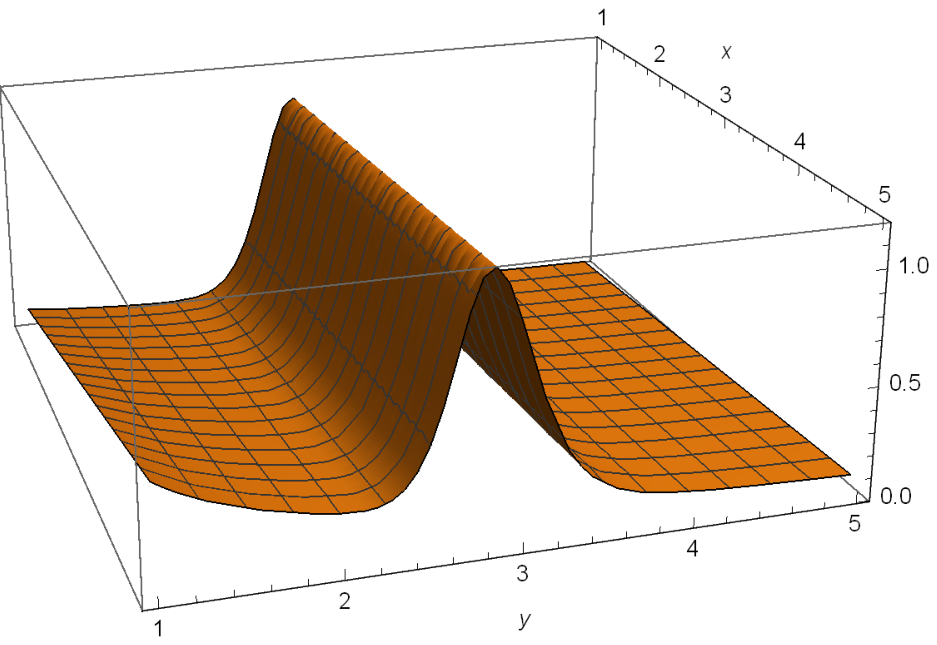}}%
\hfill
\subcaptionbox{$x=2, z = 1, t = 14$}{\includegraphics[width=0.35\textwidth]{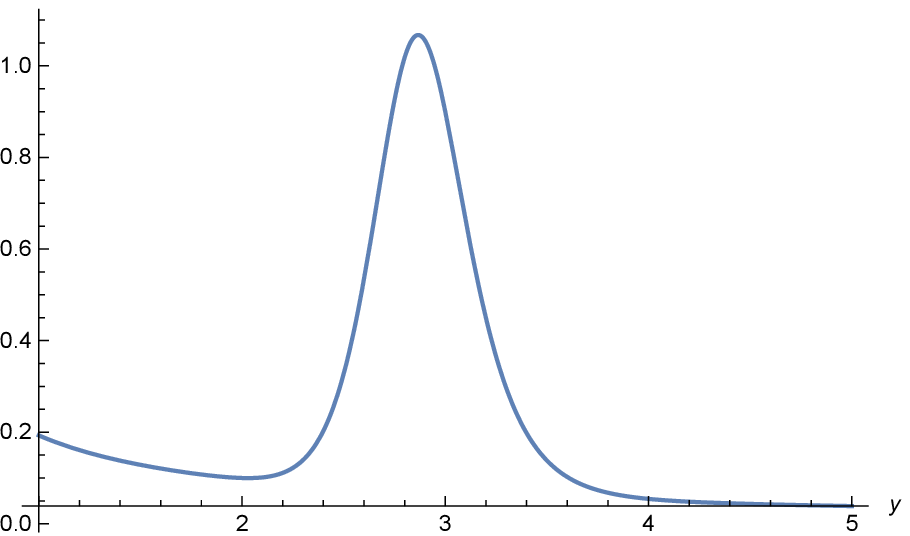}}%
\hfill 
\subcaptionbox{$z = 1, t = 14$}{\includegraphics[width=0.20\textwidth, angle=-90]{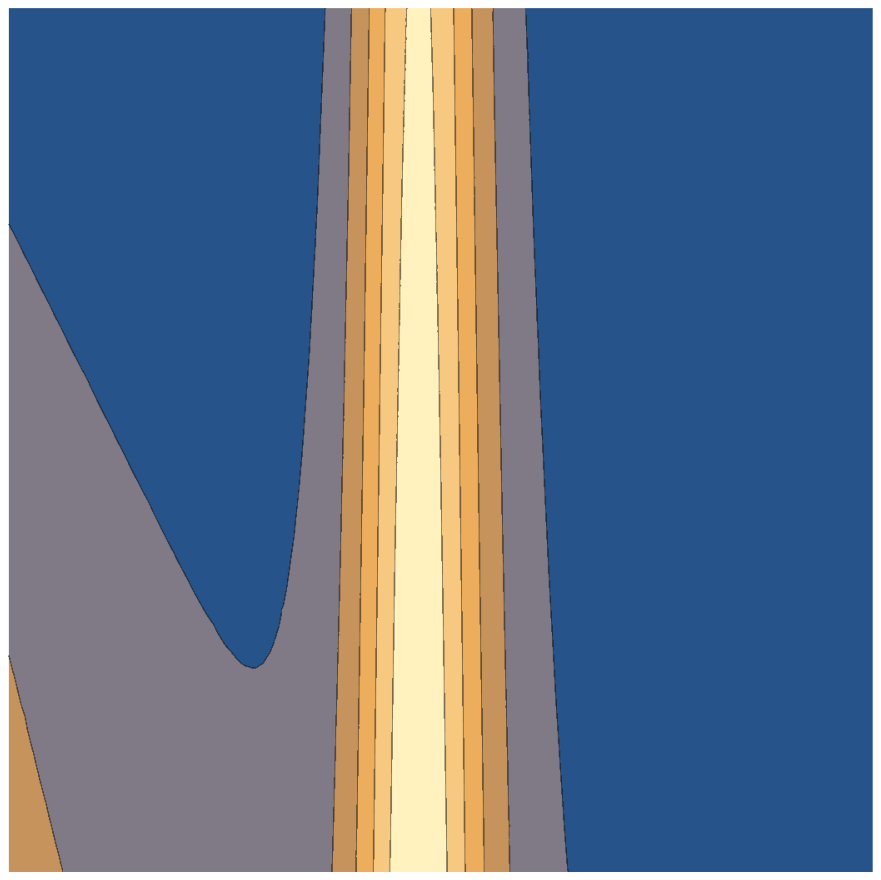}}%
\hfill 
\caption{Solitary waves solution profiles of Eq. \eqref{u16} with $f=\text{sech}\left[\frac{3 t z-5 y^2}{3y}\right]^2$.
(a) Perspective view of the real part of the single soliton solution. 
(b) The wave propagation pattern of the wave along the $y$-axis. 
(c) The corresponding contour plot. 
}
\label{f1v9b}
\end{figure}
Some particular solutions are shown in Fig. \ref{f1v9a} and Fig. \ref{f1v9b}.
\subsection{\noindent \textbf{\textit{Vector field $v_{10}$:}}}
For the infinitesimal generator 
\begin{align*}
v_{10} =  \frac{\partial}{\partial x},
\end{align*}
we have following characteristic equation
\begin{align}
\frac{dx}{1}=\frac{{dy}}{0}=\frac{{dz}}{0}=\frac{{dt}}{0}=\frac{{du}}{0}.
\end{align}
The similarity form of the solution of Eq. \eqref{kdv}
$u=F(Y,Z,T)$, with similarity variables $Y=y, Z=z$ and $T=t$.
Inserting the value of $u$  into \eqref{kdv}, we have
$F_T=0$. By solving, we get similarity solution as 
$u_{22}(x,y,z,t) = g(y,z)$, where $g(\cdot,\cdot)$ is arbitrary function.
\section{Graphical illustration of the solutions}
In this section, the numerical simulation of (3+1)-dimensional 
KdV type equation is demonstrated. The proposed method provides 
more general and plentiful new kink wave solutions with some 
free parameters. Kink waves are the type of solitary waves 
that preserve its shape when travelling down and do not 
change their shape through the propagation. Kink waves 
are travelling waves which rise or go down from one 
asymptotic state to another. Graphical representations 
of the solutions $u_3, u_4, u_5, u_{20}$ and $u_{21}$ 
are illustrated in Figures 1-8 for free choices of parameters. 
The other exact solutions could be achieved from the remaining set of solutions. 

\section{Conclusion}
This research depicts that the Lie symmetry analysis method is 
quite well-organized for extracting the exact travelling wave 
solutions of (3+1)-dimensional KdV type equation. In this paper, 
some general exact solutions in the form of kink wave, travelling wave, 
single soliton, doubly soliton, curved shaped multisoliton, explicit 
WeierstrassZeta and WeierstrassP function are constructed by 
Lie group of transformation method. The results in this article 
are excogitated and continued version of previously reported results. 
Many new exact solutions are derived, which have fruitful applications 
in different areas of mathematical physics, engineering, other fields
 of applied sciences and might provide a valuable help for researchers and physicists to study more complex nonlinear phenomena. 


\end{document}